\documentclass{ieeeaccess}
\usepackage{cite}
\usepackage{amsmath,amssymb,amsfonts}
\usepackage{algorithmic}
\usepackage{graphicx}
\usepackage{textcomp}
\usepackage{xcolor}
\usepackage{adjustbox}
\usepackage{parskip}
\usepackage{longtable}
\usepackage{needspace}
\usepackage{multirow}
\usepackage{enumitem}
\usepackage{adjustbox}

\newcommand{\papercounttotal}{270}

\begin{document}

\history{Date of publication xxxx 00, 0000, date of current version xxxx 00, 0000.}
\doi{10.1109/ACCESS.2022.0092316}

\title{Systematic Review of Extended Reality for Smart Built Environments Lighting Design Simulations}

\author{\uppercase{E. Mohammadrezaei}\authorrefmark{1}, \uppercase{S. Ghasemi}\authorrefmark{2}, \uppercase{P. Dongre}\authorrefmark{3}, \uppercase{D. Gra{\v{c}}anin}\authorrefmark{4}, (Senior Member, IEEE), and \uppercase{H. Zhang}\authorrefmark{5}}

\address[1]{Department of Computer Science, Virginia Tech, 220 Gilbert Street, Room 3213, Blacksburg, VA 24060, USA (e-mail: elliemh@vt.edu)}
\address[2]{Department of Computer Science, Virginia Tech, 220 Gilbert Street, Room 3213, Blacksburg, VA 24060, USA  (e-mail: shivagh@vt.edu)}
\address[3]{Department of Computer Science, Virginia Tech, 220 Gilbert Street, Room 3213, Blacksburg, VA 24060, USA  (e-mail: poorvesh@vt.edu)}
\address[4]{Department of Computer Science, Virginia Tech, 220 Gilbert Street, Room 3207, Blacksburg, VA 24060, USA (e-mail: gracanin@vt.edu)}
\address[5]{Department of Mechatronics
Engineering, Middle Tennessee State University, Davis Science Building (DSB), 1301 E Main St, Room 121, Murfreesboro, TN 37132, USA (e-mail: Hongbo.Zhang@mtsu.edu)}

\markboth
{Mohammadrezaei \headeretal: Systematic Review of Extended Reality for Smart Built Environments Lighting Design Simulation}
{Mohammadrezaei \headeretal: Systematic Review of Extended Reality for Smart Built Environments Lighting Design Simulation}

\corresp{Corresponding author: E. Mohammadrezaei (e-mail: elliemh@vt.edu).}

\begin{abstract}
This systematic literature review paper explores the use of extended reality {(XR)} technology for smart built environments and particularly for smart lighting systems design.
Smart lighting is a novel concept that has emerged over a decade now and is being used and tested in commercial and industrial built environments.
We used PRISMA methodology to review 270 research papers published from 1968 to 2023.
Following a discussion of historical advances and key modeling techniques, a description of lighting simulation in the context of extended reality and smart built environment is given, followed by a discussion of the current trends and challenges.
\end{abstract}


\titlepgskip=-15pt

\maketitle

\section{Introduction}
\label{sec:introduction}
 
\PARstart{T}{he} visual aspect of light has been the topic of many studies for more than 500 years.
The drawings of Leonardo da Vinci, around 1489, show the connection between the eye and the brain~\cite{DaVinci-1970-a,DaVinci-1970-b}.
In 1722 Antony van Leeuwenhoek discovered the presence of ``rod and cone cells'' in the retina~\cite{Lane-2015-a}.
In 1834, Gottfried Treviranus, a German physician~\cite{Finger-1994-a}, confirmed ``the light-sensitive photoreceptors.''
It took 150 years to find the third type of photoreceptor in the retina that responds to the light even when all input from rods and cones are blocked~\cite{Berson_2002_photo,wout_2005_visual}.

Throughout these years, many researchers and authoritative bodies worked to define standards and regulations to provide a healthy and comfortable lighting environment~\cite{dilaura2011lighting,duff20122012,RQ1_KRUITHOF}.
The foundation of these standards considers the visual aspect of the light, and most lighting designs evolved around this aspect, which includes how the eye and brain perceive light.

In recent years interest in a non-visual aspect of light is being increased.
To name a few, considering circadian rhythm in lighting design and using phototherapy for Seasonal Affective Disorder (SAD)~\cite{Partonen1998SAD} are some of the examples that have gained attention.
Human-centric lighting~\cite{houser2021human} is a term used to integrate both visual and non-visual effects of lighting to explore both the physiological and/or psychological benefits of light.
As our understanding of lighting and its effect on the environment increases, more sophisticated models and simulations are required to reflect such effects correctly.

The current advancement in XR has introduced new methods and approaches to study the effect of lighting that was never accessible to the research community.
Our goal is to gain insight into traditional simulations and recent approaches that leverage XR to provide realistic scenarios that describe some of the characteristics of light that previously were difficult to explore.
We also examine studies that explore the psychological aspect of lighting effects and the capabilities of modern lighting designs.

The remainder of the paper is organized as follows.
Section~\ref{sec:background} provides the background.
Section~\ref{sec:method} provides information about our systematic approach and  defines our research questions and approach to collecting related research.
In Section~\ref{sec:results} we provide a summary of our findings for each defined research question.
Section~\ref{sec:discussion} provides a summary of our finding and explain the necessary steps required for future studies in this area.
Section~\ref{sec:conclusion} concludes the paper.

\section{Background}
\label{sec:background}
 
Extended reality is an umbrella term that encompasses various immersive technologies, including augmented reality (AR), virtual reality (VR), and mixed reality (MR).
However, academics and professionals have been inconsistent in their use of these terms.
As a result, this inconsistency has given rise to confusion in understanding and indistinct boundaries~\cite{rauschnabel_what_2022}.
XR refers to the spectrum of experiences that combine real-world elements with digital content, creating a continuum that ranges from fully physical reality to fully virtual environments.
XR technologies have been gaining increasing attention in various fields for creating immersive, interactive and engaging user experiences.
Smart built environments (SBE) on the other hand, refer to built environments that leverage advanced technologies like Internet of Things (IoT), Artificial Intelligence (AI), and automation to optimize building performance and enhance user comfort.
In this context, XR has the potential to significantly contribute to SBE, particularly in the area of lighting design simulations.

Lighting design plays a crucial role in shaping the built environment and has a significant impact on energy consumption, user comfort, and well-being.
By using XR technologies for lighting design simulations, architects, engineers, and designers can test and visualize different lighting design options in a virtual environment.
This can help them make informed design decisions, reduce design errors, and avoid costly rework.
XR simulations can also provide a more accurate representation of the actual lighting conditions in a built environment, which is crucial for designing spaces that meet user needs and preferences.

Despite the potential benefits, research on smart lighting design within XR lacks cohesion and requires a more in-depth and structured approach.
Nonetheless, there have been instances where XR has been applied to lighting design simulations with encouraging outcomes.
For instance, some studies have used VR to create immersive simulations that allow designers to manipulate lighting conditions in a virtual environment.
This allows designers to test different lighting scenarios and quickly evaluate their impact on the space and its occupants.
Similarly, MR has been used to simulate the effect of different lighting conditions on user experience in different spaces like offices or hospitals.
By using MR, designers can observe the impact of lighting on the environment, user behavior and movement, and the effect of lighting on the overall user experience.
AR has also been harnessed in some studies to simulate lighting design and visualize lighting fixtures in real-time, which can help designers to optimize lighting conditions and identify design issues before implementation.
However, there are some challenges associated with the use of XR technologies for lighting design simulations in SBE.
These challenges include high costs associated with purchasing and maintaining XR equipment, privacy and security concerns related to collecting and using data generated by XR systems, and a lack of standards and guidelines for XR in SBE.
Despite these challenges, the potential benefits of using XR technologies for lighting design simulations in SBE are significant.
They include faster design iterations, reduced costs, improved user experience, and energy efficiency.
Therefore, a systematic review of the existing literature on the use of XR for lighting design simulations in SBE is crucial to fully understand the potential of this technology and its challenges.
This review will serve as a valuable resource for researchers, practitioners, and policymakers interested in using XR to optimize the performance and efficiency of the built environment.

\section{Materials and method}
\label{sec:method}

A systematic Literature Review (SLR) is a strategy for discovering, analyzing, and summarizing published primary research publications on a particular topic of interest.
Its goal is to answer specific research questions, identify knowledge gaps, and suggest future possibilities~\cite{rother2007systematic,xiao2019guidance}.
We reviewed the research studies and modern frameworks for SBE design and related XR technologies.
We focused on design of a smart lighting system and investigated  factors in smart lighting design in residential housing context that have a significant impact on user experience of such spaces.
There is a variety of SLR methodologies and recommendations to accomplish the SLR process, including PRISMA~\cite{selccuk2019guide} used in this study.
The three-step review phases include planning the review, conducting the review, and reporting the review.

\subsection{Stage 1: Planning the review:}

During the planning phase, a research procedure is conducted to guide the SLR process and increase the accuracy of the review process.
The research procedure is composed of the research objectives, research questions, details about the research strategy (i.e., search keywords and resources that will be searched) and eligibility criteria explaining both the inclusion and exclusion criteria of the research results.
We explain in-depth the contents of this research procedure.

\subsubsection{Research objectives}

The primary research objectives include:

\begin{description}
\item[RO1:]
\emph{Investigating current trends in smart lighting and state of the art in lighting simulation tools.}
\item[RO2:]
\emph{Previewing the current state of application of XR in smart lighting design in SBE context.}
\item[RO3:]
\emph{Investigating how modern smart lighting systems can enhance the well-being and comfort of users.}
\end{description}

\subsubsection{Research questions}

We aim to answer the following main research questions:

\begin{description}
\item[RQ1:]
\emph{What is state of the art in lighting simulation tools?}

This research question provides a focused insight into the concept of intelligent lighting design  and modern smart lighting simulation tools in the context of smart built environment.

\item[RQ2:]
\emph{How can XR technology be used for SBE lighting system modeling?}

This research question provides a comprehensive description of the concept of XR technology which is a novel concept and how it has been developed and implemented in the SBE context to assist lighting design simulation.

\item[RQ3:]
\emph{What are psychological effects of lighting on human behavior and activities?}

This research question sheds the light on the effectiveness of the light on the quality of life of people and how the different lighting design strategies can enhance this experience and their well-being and comforts.
\end{description}

To determine the scope of the SLR study, the research questions are developed using the PICOC methodology.
Population, Intervention, Comparison, Outcomes, and Context are the five criteria in the PICOC model.
The first criterion, population, is related to the investigation's participants or target system.
Interventions are actions that define the investigation's aspects and topics of interest to researchers.
The comparison refers to the aspects of study against which the intervention is evaluated.
The investigation's result is determined by the outcome.
Finally, context is related to the investigation's environment.

\subsubsection{Search strategy}

Following the identification of research objectives, we use a search technique to find studies that are related to the study's objective.
Table~\ref{tab:query} shows the primary and secondary search terms along with the final search phrase.

\paragraph{Source selection:}
Building a knowledge base from significant resources and databases is the first stage in developing a research strategy.
We incorporated four databases: ACM Digital Library, IEEE Xplore, ScienceDirect, and VT Library.

\paragraph{Search phrase:}
The process of selecting our search terms was guided by the need to address our research questions comprehensively.
We aimed to identify keywords that would best assist us in our inquiry.

To achieve this, we considered a range of terms that were relevant to our research scope, which encompasses Smart Built Environments, Extended Reality (XR), and their intersections with lighting design, simulation tools, and human factors.

The exclusion of certain similar terms was guided by our focus on relevance and the need to maintain a manageable search scope.
We aimed to keep a balance between specificity and comprehensiveness in constructing the search phrase.

The resulting search phrase, as outlined in Table~\ref{tab:query}, represents our effort to target a wide range of relevant literature while ensuring the coherence of our search strategy.
The finalized search phrases were created using logical operators AND and OR.
It combines key terms related to lighting, XR technologies, and human factors to provide a holistic perspective on the subject.

\begin{table}[tb]
\caption{Search terms.}\label{tab:query}
\scriptsize
\centering
\resizebox{\columnwidth}{!}{%
\renewcommand{\arraystretch}{1.5}
\begin{tabular}{|l|l|}
\hline
\multicolumn{1}{|l|}{\multirow{2}{*}{\textbf{Primary search terms}}} & Smart built environment, Extended reality\\ 
\multicolumn{1}{|l|}{} & Smart lighting design, Lighting simulation tools \\ \hline
\multicolumn{1}{|l|}{\multirow{4}{*}{\textbf{Secondary search terms}}} & Lighting, Modern lighting, Smart lighting design\\ 
\multicolumn{1}{|l|}{} & lighting simulation, Lighting simulation tools\\
\multicolumn{1}{|l|}{} & Lighting evaluation, Psychophysiology, Human behavior\\ 
\multicolumn{1}{|l|}{} & Human activities, Extended reality, Mixed reality \\ \hline
\multicolumn{1}{|l|}{\multirow{5}{*}{\textbf{Final search phrase}}} & (Lighting OR modern lighting OR \\
\multicolumn{1}{|l|}{} &  smart lighting design OR lighting simulation OR Lighting simulation tools \\
\multicolumn{1}{|l|}{} &  OR lighting evaluation) AND (Extended reality OR mixed reality OR  \\
\multicolumn{1}{|l|}{} &  virtual reality OR augmented reality) AND \\
\multicolumn{1}{|l|}{} &  (Human Psychophysiology OR human behavior OR human activities)\\ \hline
\end{tabular}
}
\end{table}

\subsubsection{Eligibility criteria}

Following a review of all relevant studies based on the provided search terms, a screening method was used to select only those that met the inclusion requirements.
Table~\ref{tab:eligibility} illustrates the process on inclusion/exclusion criteria.

\begin{table}[tb]
\caption{Eligibility criteria.}
\label{tab:eligibility}
\scriptsize
\centering
\resizebox{\columnwidth}{!}{%
\renewcommand{\arraystretch}{1.5}
\begin{tabular}{|l|l|}
\hline
\multicolumn{1}{|l|}{\textbf{Inclusion criteria}} & \textbf{Exclusion criteria} \\ \hline
\multicolumn{1}{|l|}{\multirow{1}{*}{Publication Year: 1994--2023}} & Document is not related to the search topic.\\ \hline
\multicolumn{1}{|l|}{\multirow{2}{*}{Peer reviewed documents}} & Workshop, panel, tutorial, seminar, interview, \\ 
\multicolumn{1}{|l|}{} & or poster.\\ \hline
\multicolumn{1}{|l|}{\multirow{1}{*}{Available online}} & Document is book or magazine.\\ \hline
\multicolumn{1}{|l|}{\multirow{1}{*}{Document responds to at least one of the RQs.}} & Marginal publication venue. \\ \hline
\end{tabular} 
}
\end{table}

\subsection{Stage 2: Conducting the Review}

The reviewing process was conducted through three main phases:

\subsubsection{Phase-1}

Throughout this stage, we evaluated the papers' relevance to our study based on their title, abstract, introduction, and conclusion.
Throughout this process, we sought to be as inclusive as possible to ensure that no research were overlooked.
This phase led to the exclusion of 182 studies.

\subsubsection{Phase-2}

We examined the trustworthiness of the selected articles using two credibility criteria in the second step.
Once a study meets one of these two credible requirements, it is included:
\begin{enumerate}
\item 
The research should be published at a high-impact and well-known conference or journal.
As a result, we established criteria to include journal articles with an impact factor of at least 3.
The inclusion of conferences was based on their repute.
\item
Studies having more than 10 citations will be maintained and carried on to the next round.
As is well known, the quantity of citations represents the study's importance and significant impact.
114 studies were eliminated at the end of this stage.
\end{enumerate}

\subsubsection{Phase-3}

\begin{figure}[t]
\centering
\includegraphics[width=0.925\linewidth]{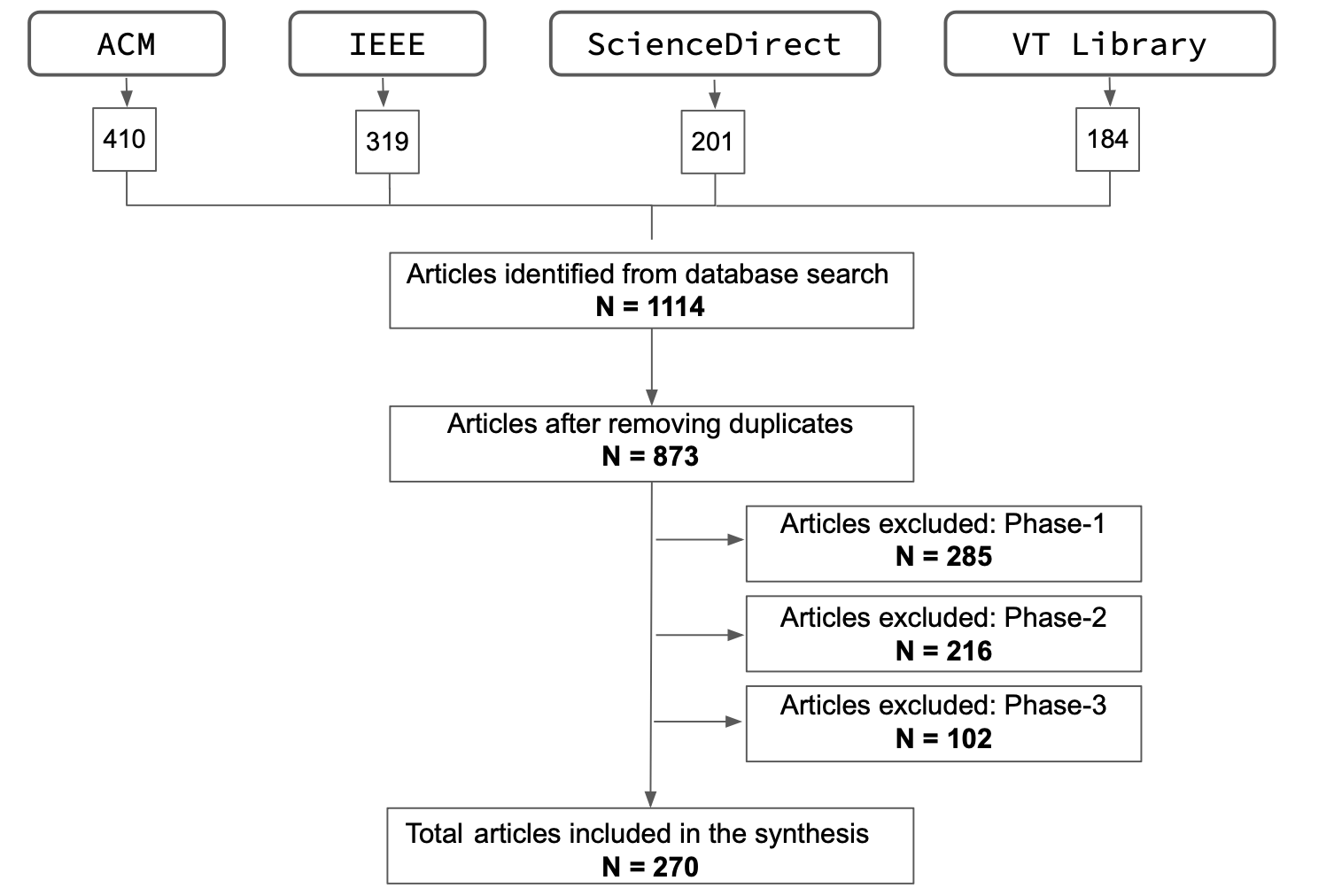}
\caption{Flowchart demonstrating method of screening phases.}
\label{fig:phaseA}
\end{figure}

The final phase is a detailed screening of the research that have been submitted.
The entire texts of these papers were examined using the eligibility criteria, and a total of \papercounttotal\ publications were found to be relevant to our study topics to some extent and were chosen as primary studies.
Figure~\ref{fig:phaseA} is the illustrated result of the complete screening process.

\subsection{Stage 3: Reporting the review:}

A data extraction approach was used to relocate the included primary studies.
Each primary study was downloaded from its digital repository and given a unique file ID in this strategy.
To fill in the details of each study, an excel spreadsheet was prepared.
The study's abstract, country of origin, user experience problems, set of factors, significance and limitations of the study, study gap, recommended solution, and future work were all retrieved and placed in their appropriate columns.
As a result, as shown in the next section, the saved analysis information was used to analyze the data and respond to the RQs.

\section{Results}
\label{sec:results}

\subsection{RQ1: What is the state of the art in lighting simulation tools?}

The field of smart lighting design is experiencing rapid growth and is predominantly utilized in disciplines such as architecture and entertainment.
Unlike traditional lighting design, smart lighting design brings new functionalities to traditionally lit spaces.
These include the ability to influence the atmosphere, improve energy efficiency, highlight areas with higher foot traffic, and preserve lighting in unpopulated regions~\cite{E.Mohammadrezaei}.

Due to the rigorous requirements for reality representation, lighting simulation is complex, but it also offers various levels of complexity for multiple users within the same field.
One of its many advantages is that it can give researchers quicker and more efficient ways to compare more complex data that would have taken much longer.
It also takes the place of rigid scale models and large, energy-intensive equipment.
Designers can compare and change the effects of artificial and natural lighting on project alternatives.
It is simpler to picture the advantages of well-designed natural and artificial lighting.
Integrating lighting simulation with other building simulations and the design process is possible.

A critical literature review helps to advance various ideas and map out field developments.
This is significant since lighting simulations are rapidly replacing traditional verification methodologies.
In a poll conducted in 1994, about 77\% of respondents said they employed both computers and actual models in their professional work.
Participants who did not use any program for daylight prediction fell to 21\% by 2004~\cite{reinhart2006findings}.
As architecture and engineering students become more accustomed to computer modeling, this number may drop even more~\cite{pentilla2003architectural}.
The use of lighting simulation will also rise due to more sophisticated ways to prove compliance with newer, complex construction rules and certifications~\cite{de2009preconditions,ibarra2009daylight}.

This section contains in-depth information and a discussion of the many simulation technologies that various researchers have used to simulate their modules and prototypes.
Based on this analysis, extensive data that includes all of the simulation tools used for artificial lighting in buildings and the many factors that go along with them have been summarized in Tables~\ref{tab:simpleSim} and~\ref{tab:integratedSim}.
The primary goal of this study is to identify the simulation tool that is most frequently used to simulate illumination in building envelopes.

Throughout this review, a total of 65 papers that were published between the years 2001 and 2017 were reviewed.
Only 25.71\% of conference papers, which were highly relevant to the construction and energy sectors, as shown in Figure~\ref{fig:phaseA}, were considered for this review, accounting for nearly 74.29\% of peer-reviewed articles.
In 2013, 15.71\% of the highest publishing ratio was noted.
According to the current literature analysis, 14.28\% in 2016, 10\% in 2012, and just 8.57\% in 2017 have been published as of this writing.

\subsubsection{Historical Developments}

Computer calculations are ideal for predicting light behavior.
Formulas claim it would take a long time to solve them manually.
As the complexity of the scene being calculated increases, so does the complexity.
The design of electric lighting installations in sealed rectangular rooms was one of the earliest repeated applications to use computer calculation—some of the earliest initiatives in that direction date back to the 1970s~\cite{hirata1999improvement}.
During the same decade, efforts were undertaken to use streamlined methods to incorporate calculations for both natural and artificial illumination under set conditions~\cite{plant1973computer}.
The output was mainly numerical.

Computer graphics simultaneously made significant advancements, leading to notable results in the 1980s.
Using computational methods, researchers enhanced the calculation and representation of light falling on arbitrary volumes~\cite{nakamae1995photorealism}.
These estimations were not limited to any particular scenario because they were based on light travel and material characteristics.
Results could be retrieved as numbers or as photos that closely resemble reality.
Physically-based rendering was established as a distinct computational field~\cite{kniss2003model}.
The fields of lighting research, architectural representation, and cinematography are some examples of those who would directly profit from these advancements.
Entirely correct computations, however, turned out to be challenging to specify, computer-intensive, and subject to implicit constraints~\cite{wilkie2021predictive}.
Early lighting modeling algorithms were thus constrained to simplistic geometries, no consideration of sunshine or energy, and erroneous output of only a few parameters~\cite{svendenius1995searching}.
Modern models address these characteristics; however, developing computation-efficient techniques to solve the ``global illumination problem'' is still a work in progress~\cite{ulbricht2006verification}.

Even while advancements in the two primary areas contribute to the other, lighting simulation can still be split into these categories.
The first involves creating artistic visuals using photo-realistic rendering.
Physically-based visualization is the second area and the subject of this review (also known as predictive rendering).
It is concerned with accurately depicting and predicting reality under predetermined circumstances while adhering to physical rules~\cite{ward1998rendering,Fmoeck1996computer}.

\subsubsection{Modelling}

A precise and definite set of instructions is an algorithm.
They are used in simulations to address issues with lighting distribution to combine several physical formulas~\cite{ochoastate}.
Dutre et al. thoroughly described the foundations and advancements of lighting simulation methods.
They claim that current physics theories (such as quantum optics) explaining how light moves through all kinds of media are too intricate for image production and computation.
Instead, physical formulas are known from a simplified geometrical optics model and energy conservation.
This model can produce numerical results for the simulation space and handle most illumination issues with various light sources.
However, when diffusing or reflecting media are involved, challenges arise (such as advanced optical materials)~\cite{dutre2006advanced}.
Additionally, steady-state light distribution is assumed.
These limitations may explain discrepancies between observations and simulations in specific modeling scenarios.

The geometrical optics model was gradually implemented to address the global illumination issue.
Hurdles were posed by calculating surface reflectivity according to the material.
Different sub-models that detail their bidirectional reflection distribution functions can be used to depict them (BRDFs).
The first steps were taken in computer graphics, but the results were only able to assign colors (rasterization) and determine hard shadows~\cite{pineda1988parallel}.
Raytracing and radiosity, two of the most widely used illumination techniques today, were later developed.

\paragraph{Lighting simulation algorithms and calculation tools}
Researchers provided an early assessment of daylighting simulation resources and algorithms as an inventory of approaches for comprehensive analysis tools.
Currently, there are three different categories of lighting simulation methods.
These are scene-dependent algorithms, view-dependent algorithms, and direct calculations.
Modern models can use one of them or a combination of them.
Although some phenomena (such as fog, specific diffusing materials, etc.) are challenging to calculate using these methods alone, some approximations reflect their effects~\cite{dutre2006advanced, kniss2003model}.

\textbf{Direct calculations:}
They are used for lighting directly from light sources.
The sun, daylight apertures (which include both the sun and the sky), and luminaires are some examples of these sources.
The phrase is primarily used in the text to link models that involve the design and calculation of luminaires.
They entail particular physical simplifications and formulations.
They cover the majority of ordinary circumstances and are frequently defined by national standards.
The lumen approach is one prominent illustration.

\textbf{View-dependent algorithms:}
Algorithms like raytracing, are categorized based on the angle from which the model computes rays in a scene.
Rays are traced either forward from the light source, backward from the observer's eyes, or both ways~\cite{lafortune1993bi}.
It is an image-bound approach as a result.
It offers a method for calculating light phenomena resulting from direct illumination, specular surfaces, and reflections.
When computing several diffuse reflections or indirect lighting effects, there are, nevertheless, some restrictions.

\textbf{Scene-dependant algorithms:}
Algorithms like radiosity were adopted from heat transfer methods in order to simulate illumination, scene-specific procedures~\cite{willmott1997empirical}.
They benefit from a scene-based approach.
Surface elements can be used to categorize the scene. Each surface's radiometric values are calculated without regard to the view.
Its inability to effectively handle specular reflections is one of its flaws.
Surface meshes were used to introduce model refinements.
Umbras and penumbras can be calculated in this way~\cite{wang1992new}.
The method is mainly utilized for lighting calculations rather than image production because the calculation methods are more complicated.
Innovations are being made to lighten the load on hardware and increase its effectiveness~\cite{wang2009new}.

\textbf{Calculation tools:}
Statistical sampling is necessary to collect values within an acceptable time frame for implementing these algorithms on current computing systems.
There are deterministic approaches to conventional numerical techniques.
However, the most popular methods are called Monte Carlo methods.
They are well-liked because the methodology presupposes that the sample's predicted value is the actual value for that sample.
Once estimates have been located, the problem can be solved by averaging the estimates.
This can produce precise findings, but the algorithms need to run long enough to collect many samples.
Correctors must be included since some areas can be left with no values (referred to as ``noise'').
Although they have accuracy limitations, Monte Carlo approaches are sometimes the best option for solving some physical issues.
A summary on lighting simulation algorithms is given in Table~\ref{tab:algorithm}.

\begin{table}[tb]
\caption{Lighting simulation algorithms currently available.}
\label{tab:algorithm}
\scriptsize
\centering
\resizebox{\columnwidth}{!}{%
\renewcommand{\arraystretch}{1.5}
\begin{tabular}{|l|l|}
\hline
\multicolumn{1}{|l|}{\textbf{Algorithm}} & \textbf{Value} \\ \hline
\multicolumn{1}{|l|}{\multirow{3}{*}{View-dependent}} & Forward ray tracing \\ 
\multicolumn{1}{|l|}{} & Backward ray tracing \\ 
\multicolumn{1}{|l|}{} & Bi-directional ray tracing \\ \hline
\multicolumn{1}{|l|}{\multirow{4}{*}{Scene-dependent}} & Radiosity \\ 
\multicolumn{1}{|l|}{} & Photon map \\ 
\multicolumn{1}{|l|}{} & Integrative approaches \\ 
\multicolumn{1}{|l|}{} & Multi-pass approaches \\ \hline
\multicolumn{1}{|l|}{\multirow{2}{*}{Direct Calculation}} & \multirow{2}{*}{For artificial lighting follows national standards} \\
\multicolumn{1}{|l|}{} &  \\ \hline
\multicolumn{1}{|l|}{\multirow{2}{*}{Calculation Aids}} & Deterministic methods (Classical approaches) \\ 
\multicolumn{1}{|l|}{} & Statistical sampling methods (Monte   Carlo) \\ \hline
\end{tabular}
}
\end{table}

\begin{figure}[t]
\centering
\includegraphics[width=0.925\linewidth]{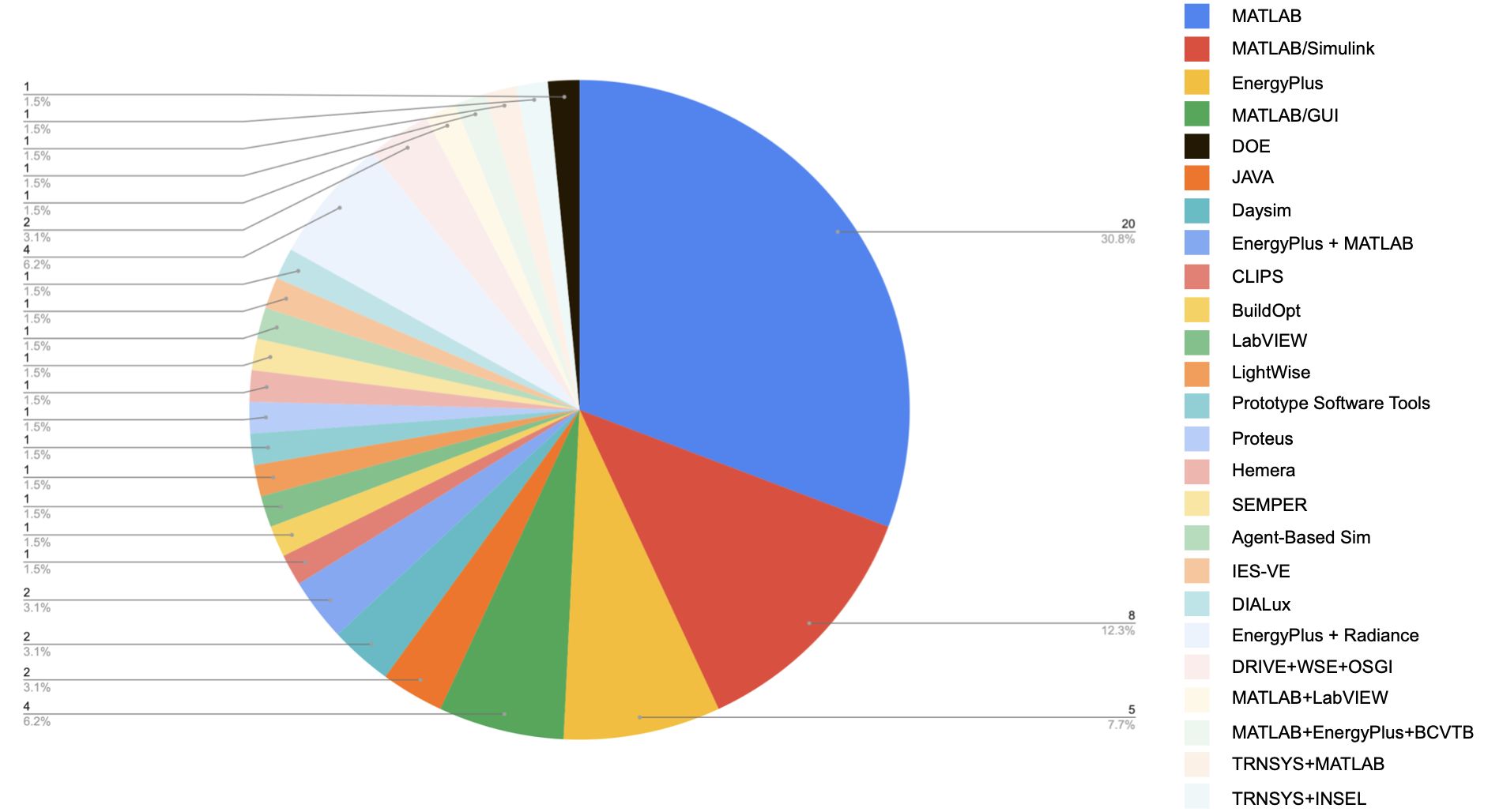}
\caption{Lighting simulation tools publications (65).}
\label{fig:Percentage}
\end{figure}

\paragraph{Lighting Simulation tools}
This review includes 65 specialized research publications that apply simulation to building lighting.
The purpose of the survey, which focused on learning about the simulation technologies used for lighting simulations, was to thoroughly examine these 65 publications.
According to the study, the researchers employed both single tools and, in some cases, the combination of two simulation tools (Integrated tools) to simulate illumination.
MATLAB (Matrix Laboratory), MATLAB-GUI (Graphical User Interface), EnergyPlus, Design of Experiments (DOE), JAVA, Laboratory Virtual Instrument Engineering Workbench (LabVIEW), Daylight Simulation (DaySIM), C Language Integrated Production System (CLIPS), and Building Optimisation are the singly used simulation tools (BuildOpt).
In addition, EnergyPlus/Radiance, EnergyPlus/MATLAB/LABVIEW, MATLAB/Transient System Simulation (TRNSYS), and TRNSYS/Integrated Simulation Environment Language (INSEL) are integrated simulation tools utilized for the simulation.
Figure~\ref{fig:Percentage} depicts the information retrieval for the frequency of simulation tools.
It is percentage representation of the combined data highlights the prevalence of the simulation techniques examined in the literature.

\paragraph{Simple simulation tools}
Twenty (30.8\%) of 65 publications used MATLAB to simulate the prototypes.
A fourth-generation, multi-paradigm programming language for numerical computing is called MATLAB.

The power needed by environmental parameters to maintain a comfortable indoor environment with the control of associated actuators was calculated by Shaikh et al.~\cite{matlab_shaikh2016intelligent} using a fuzzy controller.
The authors asserted a 31.6\% energy efficiency with an 8.1\% increase in comfort score.
By applying the enhanced version of their earlier work, the optimization toolbox, Shaikh et al.~\cite{matlab_shaikh2013indoor} claimed to save 32.69\% of energy.
They used the Multi-Objective Optimisation Genetic Algorithm (MOGA) algorithm to create the Pareto front trade-off solution set.

To estimate the system's output signals that use appropriate fuzzy sets represented by a rule set database and membership function, Rodriguez et al.~\cite{matlab_rodriguez2015fuzzy} employed the MATLAB fuzzy inference algorithm.
The findings demonstrated that the suggested system was able to maintain the required visual comfort levels using the available resources while at the same time using less artificial light, which reduced power consumption.
Pennacchio and Wilkins~\cite{matlab_penacchio2015visual} employed the MATLAB closest neighbor approach for computational measurement.
Images were cropped and enlarged to $256 \times 256$ to save the central square region, and a 2D Fourier amplitude spectrum was produced for each image.

In a different study, Shaikh et al.~\cite{matlab_shaikh2014stochastic} used fuzzy controllers to compute the power requirement for regulating the appropriate subsystem actuator to maintain the desired comfort.
By enhancing the building intelligence, the proposed Genetic Algorithm (GA) can balance the overall power consumption and the user's convenience.
As a result, far less power is now required.
The computation of the power consumption for the artificial illumination control system was also created by Shaikh et al.~\cite{matlab_shaikh2013building} utilizing the MATLAB fuzzy inference system (FIS) controller.
The implications of various membership functions have also been used to validate the inference models for each agent.
Further described by Shaikh et al.~\cite{matlab_shaikh2013indoor} was a fuzzy inference method for controlling actuator systems for visual comfort in addition to other comfort parameters and power consumption.
The Mamdani implication method was used with approximate rule-based reasoning and the widely utilized centroid defuzzified way.
For the actuator to operate and to maintain the desired level of indoor illumination comfort, required power and available power were synchronized.
According to the modeling results, the necessary power has fallen linearly.

The fuzzy control system was introduced by Kristl et al.~\cite{matlab_kristl2008fuzzy} to create a comfortable environment inside the structure.
A secondary PID controller and the primary fuzzy controller made up the control loop structure of the illumination level.
The fuzzy control system sets the placement of the shading device and rendering parameters for indoor and outdoor environments.
Shaikh et al.~\cite{matlab_shaikh2013robust} created the fuzzy-based model with a membership function and employed the user-selected set points as inputs.
At the same time, the fuzzy controller used external environmental factors.
The model was subjected to the proportional control rule base to provide a discrete output.
The retrieved fuzzy control models include models for both comfort and power usage.

The fuzzy control system architecture (FCSA) was also presented by Shaikh et al.~\cite{matlab_shaikh2013building} for resolving conflicts between energy usage and occupant comfort.
The control for visual comfort and other comfort factors has been designed using FIS.

The Multi-Island Genetic Algorithm (MIGA) was devised by Ali and Kim~\cite{matlab_ali2013energy} to maximize occupant comfort while minimizing power usage.
The error difference between the user's input parameters and the actual environment served as the fuzzy logic's input.
By reducing the error difference with MIGA/GA, the trade-off between comfort index and energy saving was achieved.
Wang et al.'s~\cite{matlab_wang2012multi} proposal used model-based information fusion and the Ordered Weighted Averaging (OWA) aggregation method to control the energy and comfort of indoor building environments.
Fuzzy controls and the power necessary to maintain the highest level of comfort were used.

This method resulted in a 3\% increase in overall comfort and a 9\% reduction in energy use.
Wang and Wang~\cite{matlab_wang2012occupancy} employed a Particle Swarm Optimization (PSO) algorithm to achieve comfort and energy efficiency in a building environment simultaneously.
PSO is a computational technique that iteratively optimizes issues.
Utilizing the suggested local illumination controller agent can save about 31\% energy.
In a different work, Z.
Wang et al.~\cite{matlab_wang2012intelligent} used PSO to improve the entire system's intelligence when combined with the building and microgrid systems.
Residents have access to a GUI-based portal where they can input their preferences and track the outcomes.

According to the measured indoor illuminance, Lah et al.'s fuzzy controller~\cite{matlab_lah2006daylight} was presented for the placement and decision-making of the roller blind to keep the illumination within the chamber at the desired level.
Depending on the sky circumstances, maintaining the appropriate interior illumination (between 750 and 1500 lx) with roller blind movement results in an interior luminous efficacy ranging from 2 to 10 lm/W.
The only exceptions are in the mornings and evenings when the sun shines brightly, and the roller blind is open 90\% or more.
Rutishauser and Joller proposed the fuzzy rule base, and Douglas~\cite{matlab_rutishauser2004control} evaluated system performance.
The suggested fuzzy online system used a real-time unsupervised learning technique to build a fuzzy rule base from extremely sparse input in a non-stationary environment.
There are two types of rules: static and dynamic.
Static rules set the system's fixed bounds but are active.

For changeable transparent aspects of the building envelope, Lah et al.'s~\cite{matlab_lah2005fuzzy} proposal of fuzzy rule-based elements was made once more.
In their pre-simulation of the louver angle, Park, Augenbroe, and Messadi~\cite{matlab_park2003daylighting} used RADIANCE and one of MATLAB's constrained nonlinear optimization algorithms to ``identify a minimum of a constrained nonlinear multivariable function (FMINCON).''
To create a wisely calibrated fuzzy logic controller (FLC) to control HVAC systems in terms of energy performance and indoor comfort requirements, Alcala et al.~\cite{matlab_alcala2003fuzzy} considered using a genetic algorithm.
To create a unique adaption system based on user preferences for blind position learning, Guillemin and Molteni~\cite{matlab_guillemin2002energy} used a genetic algorithm.
The components that made up the proposed adaptation process were a pre-processing wish module, a GA engine module, and a sensitivity filter.

\paragraph{Integrated simulation tools}
Many researchers have embraced integrated simulation technologies to simulate their prototypes.
Several of these crucial tools are covered below:

\textbf{EnergyPlus and Radiance:}
Four out of 65 publications (6.2\%) describe the use of the EnergyPlus and Radiance software for four cities with various climates—Montreal, Canada; Boulder, USA; Miami, USA; and Santiago, Chile—Bustamante et al.~\cite{energy_rad_bustamante2017integrated} examined two external and moveable CFSs.
To ensure visual comfort and reduce overall energy usage, the maximum allowed incident irradiance has to be determined.
To do this, mkSchedule was used to create a timetable for the luminaires' power fraction and CFS position~\cite{energy_rad_vera2016flexible}.
EnergyPlus~\cite{ehrlich2001simulating} was used for the energy simulations, and Radiance~\cite{ward1994radiance} was used for the illumination simulations.
Vera et al. optimized the fixed exterior complex fenestration systems (CFS) of four separate offices—Montreal (Canada), Boulder (USA), Miami (USA), and Santiago (Chile)—located in various climatic regions.
The total energy consumption and two visual comfort criteria—annual sunlight exposure (ASE) and spatial daylight autonomy (SDA)—were pursued through an optimization problem.
The GenOpt engine was paired with the hybrid PSOHook Jeeves (HJ) algorithm, mkSchedule, Radiance, and EnergyPlus to perform integrated lighting and temperature calculations.

Vera et al.~\cite{energy_rad_vera2016flexible} presented the mkSchedule, which includes the three-phase method and uses scripts built in the Lua programming language for the flexible and prompt production of a schedule for the CFS position and luminaries' power fraction.
Three case studies from a San Francisco (CA, USA) workplace were used to apply the established technique to show its adaptability and time-saving capabilities.
For artificial illumination, on/off and dimmable control schemes were investigated.
The findings demonstrated that the CFS's position and luminaries' power fractions responded appropriately to the control algorithms of the CFS and luminaries.
The schedules were successfully constructed by mkSchedule and then imported to EnergyPlus and Radiance to acquire the annual solar heat gains and illuminance.
Huang and Niu used integrated simulation platforms to research a numerically based silica aerogel glazing system~\cite{energy_rad_huang2015energy}.
Investigated were the energy consumption and visual comfort of glazing systems in commercial building envelopes.
EnergyPlus was used to test the glazing system's energy efficiency.
In contrast, the Radiance tool measured the visual comfort index.

\textbf{EnergyPlus and MATLAB:}
Two out of 65 publications (3.1\%) combined the usage of MATLAB and EnergyPlus.
A non-dominated sorting genetic algorithm (NSGA-II) code was built using MATLAB for the simulation-based optimization process by Delgarm et al.~\cite{energy_mat_kim2016simulation} utilizing EnergyPlus software to conduct the simulation.
An Energy Management Control (EMC) system that incorporates predictive dynamic building models, day-ahead forecasts of load, energy price, and occupancy needs, and supports micro-zoning-based active occupant-driven control was proposed by Ji, Lu, and Song~\cite{energy_mat_ji2012energy}.
The EMC functions as a high-level strategy planner that enables the building automation system (BAS) to utilize all available information for optimal planning and operation.
It is based on a dual-loop control design and implemented in the outer loop.

\textbf{MATLAB and LabVIEW:}
One publication out of 65 (1.5\%) combined the usage of LabVIEW and MATLAB technologies.
Xiong and Tzempelikos~\cite{mat_lab_xiong2016model} combined MATLAB and LabVIEW for their control platform.
Since data measurements and controls were handled through LabVIEW and the daylighting-glare model was created in MATLAB, the two simulation platforms each offer unique performance characteristics.
The LabVIEW tool's MathScript RT Module provided integration inputs from the MATLAB model.
Consequently, the models are run in real-time, and the outcomes are collected using both tools.

\textbf{MATLAB, EenergyPlus, Building Controls Virtual Test Bed (BCVTB):}
In one of  65  publications (1.5\%), the proposed system was simulated using MATLAB, EenergyPlus, and BCVTB.
The energy and visual comfort of lighting and daylighting management systems were analyzed by Shen, Hu, and Patel~\cite{mat_ene_bc_shen2014energy}.
MATLAB, EnergyPlus, and BCVTB were used to model the proposed approach.
To conduct simulations for building energy management systems, users of BCVTB can pair a variety of simulation programs and control algorithms.
EnergyPlus provided data on energy and daylight generation and the building model.
In contrast, MATLAB provided the ability to develop blind control algorithms with dynamic statistical occupancy.

\textbf{TRNSYS and MATLAB:}
One out of 65 publications (1.5\%) employed the TRNSYS and MATLAB tools for the illumination simulation.
Dounis and Caraiscos~\cite{dounis2007intelligent} have merged TRNSYS and MATLAB tools.
The control system was implemented in MATLAB, and the model acted as the existing system implemented in TRNSYS.
The simulation results demonstrated that the suggested intelligent control system successfully balanced energy conservation and occupant comfort.

\textbf{TRNSYS and INSEL:}
One out of 65 publications (1.5\%) combined the usage of the TRNSYS and INSEL tools.
These tools were frequently used by Santos et al.~\cite{trn_ins_colmenar2013solutions} to design and develop HVAC systems and construct rooms in the simulation mode.
These tools enable project planning for the ideal indoor lighting design.

\textbf{DRIVE, WSE and OSGi:}
The integration of DRIVE (Distributed Responsive Infrastructure Virtualization Environment), WSE (WebSphere Sensor Events), and OSGi (Open Services Gateway Initiative) was employed for system simulation in one out of 65 publications (1.5\%).
Using a research tool named DRIVE~\cite{chen2008drive}, H. Chen et al.~\cite{dr_wse_osgi_chen2009design} created event processing logic as components and flow diagrams.
Runtime artifacts produced by this software comprise configured scripts and codes that can be used with WSE and controlled by OSGi platforms.

\subsubsection{Evaluation}

We provide an in-depth discussion of the many simulation tools that various researchers have used to simulate their modules and prototypes.
Based on this analysis, comprehensive data that includes all of the simulation tools used for artificial lighting in buildings and the many factors that go along with them have been summarized in Table~\ref{tab:simpleSim2} and~\ref{tab:integratedSim}.
The primary goal of this analysis is to identify the simulation tool that is most frequently used to simulate illumination in building envelopes.

\paragraph{Accuracy validations and model comparisons}
Comparing simulation models is a logical conclusion when there are numerous of them available.
Comparisons based on accurately simulating the built environment and comparisons made in controlled laboratory environments make up the two main categories of evaluations discussed in the literature.
There are benefits and drawbacks to each strategy.
Due to inherent methodological restrictions, as explained in the last section, comparing findings acquired using different methods is challenging.

\textbf{Validations with built realities:}
Most comparisons in this category look at lighting simulation tools architects can use to create and assess designs.
Roy provided a classical, thorough analysis of the programs accessible at the time.
The ``perfect rendering package'' for usage in architecture was compared with functionality, usability, and application to design settings~\cite{roy2000comparative}.
Christakou and Amorim carried out a similar study in 2005 for the Brazilian context~\cite{christakou2005daylighting}.
    
Four simulation programs were also evaluated by Ashmore and Richens~\cite{ashmore2001computer}, ADELINE- Radiance and three programs now unsupported: Lightscape, RadioRay, and Microstation 7).
As a standard, two artificially overcast skies over a scale model were used.
Using reflectances from the model, programs had to simulate measurements of a small room next to a huge courtyard.
The results from simulation programs were decent overall, but poor near the window.
The size model and two artificial sky that these writers chose were acknowledged as mistake sources.
    
Ng discovered that selecting the appropriate reflectances impacts simulation outcomes.
Heavily occluded locations in Hong Kong were examined using Desktop Radiance and Lightscape.
Measurements inside test environments were compared to computer simulations.
The sky model was cloudy and created using local meteorological data.
There were noted large relative inaccuracies~\cite{ng2001study}.
Mardaljevic attributed these mistakes to the substantial role of the reflectance estimation of nearby surfaces~\cite{mardaljevic2004verification}.
A model of a small Singapore museum in cloudy weather was created by Ng et al. in 2001.
Then, modeling nearly matched daylight readings.
According to Li et al., furniture reflectances in a Hong Kong classroom also impacted accuracy findings~\cite{li2004predicting}.
    
The aforementioned illustrations of model validations show how they were utilized in various scenarios faced by building designers and the kinds of data they use.
Additionally, they emphasize the need for models to adequately depict reality while simultaneously accommodating user expectations for program manageability.

\textbf{Validations in controlled lab environments:}
In most cases, comparisons under controlled conditions were conducted to show model accuracy in predicting illumination data.
    
Under IEA-SHC Task 21, Fontoynont et al. conducted the first set of standardized assessments for simulation tools.
Testing of the tools was done in an atrium with various surface finishes.
Scale models served as benchmarks.
It was discovered that input data quality had an impact.
Accuracy was affected by careful modification of simulation parameters~\cite{fontoynont1999validation}.
At the time, time-static models were found to exist.
Maamari and Fontoynont carried on Task 21's work by creating several test scenarios~\cite{maamari2003analytical}.
The CIE test scenarios would be based on these.
Two light simulation tools—one for general rendering and the other for designing electric lighting—were contrasted to show how they differ~\cite{maamari2006experimental}.
    
To create a dynamic simulation method, Reinhart and Walkenhorst did Radiance validation of a considerable number of sky situations~\cite{reinhart2001validation}.
A specific full-scale model with external blinds was the subject of this examination.
Reinhart tested standard equipment like shelves and curved curtains to validate Radiance and 3D Studio (mental ray photon map)~\cite{reinhart2009experimental}.
    
Developers can test algorithms under less unpredictable scenarios by validating models under controlled circumstances.
They should be carried out in the preliminary stage before being incorporated into a complete simulation model.

\textbf{Standard comparison techniques:}
By comparing lighting simulation tools, it may appear straightforward to rank or select among them.
In actuality, a variety of factors influence the result.
According to Maamari and Fontoynont, replicating built realities or scale models introduces uncertainty~\cite{maamari2003analytical}.
These consist of measurements, geometry, and the material's photometric qualities (such as lamp depreciation and surface reflectivity).
Most areas worldwide do not have access to hourly data of good quality for daylighting simulations~\cite{david2010method}.
On the other hand, while evaluating computer programs in a pure laboratory setting, routine construction activity is disregarded~\cite{galasiu2002applicability}.
    
There have been suggestions to standardize comparison techniques to avoid the challenges mentioned above.
The Cornell box was one of the first methods used to test the early rendering algorithms~\cite{goral1984modeling}.
The method's primary use for visual comparison is a drawback~\cite{ulbricht2006verification}.
BRE proposed a dataset focused mainly on daylight performance in particular circumstances~\cite{mardaljevic2001bre}.
A set of tests with only artificial lighting were proposed by CIBSE~\cite{slater2002benchmarking}.
Based on work by~\cite{maamari2003analytical,maamari2006experimental}, the CIE recommended a set of test cases to be used as benchmarks~\cite{cie2006tc}.
These test cases cover as many variants as possible while keeping everything straightforward and repeatable.
Both artificial and natural light and various kinds of surfaces are present in the cases.
    
The validation techniques discussed, together with their drawbacks and accuracy, are summarized in Table~\ref{tab:val_table}.

\begin{table}[tb]
\caption{Lighting simulation validation techniques}
\label{tab:val_table}
\scriptsize
\centering
\begin{tabular}{|p{0.5in}|p{0.75in}|p{1in}|p{0.4in}|}
\hline
\textbf{Method} & \textbf{Advantages} & \textbf{Disadvantages} & \textbf{Accuracy} \\ \hline
\multirow{3}{0pt}{Validation with built realities} & \multirow{3}{60pt}{Tests on buildings, live measurement} & Hard to replicate somewhere else & \multirow{3}{*}{5--20\%} \\ 
 &  & Hard to replicate sky conditions &  \\ 
 &  & Hard to replicate reflections &  \\ \hline
\multirow{2}{0pt}{Laboratory tests} & Controlled settings & Detaches from model usability & \multirow{2}{*}{2--7\%} \\ 
 & Element testing & Only possible with specialized tools and data &  \\ \hline
\end{tabular}
\end{table}

\paragraph{Methods for Solving Various Lighting Simulation scenarios}
Certain issues can be resolved differently depending on the assumptions used by current illumination simulation techniques about steady-state and geometrical distribution.
They include modeling of new, untested components, numerous performance simulations, and time-dependent simulations (to analyze user behavior and responsive elements) (prototyping).
To duplicate effects over a period of time, time-span calculations need to be broken up into very small chunks.
Consequently, a different strategy is needed.

\textbf{Dynamic vs. static:}
Simulation algorithms and tools presuppose a steady-state light distribution.
For a lot of situations, this assumption may be accurate.
Problems involving daylight must take into account light's dynamic behavior.
Annual irradiation studies, occupant reactions to variations in light quality, and dynamic shadow components are a few examples.

Typically, the time period is broken up into manageable data chunks.
On these segments, the model runs simulations and compiles the findings.
This is done by obtaining information about sun radiation from local weather databases (usually hourly averages).
As permitted by the weather file, interpolations anticipating sun position and radiation must be done for fewer time increments~\cite{walkenhorst2002dynamic}.
Radmap and GenCumulativeSky are examples of tools that produce cumulative sun radiation maps using such ideas~\cite{robinson2004irradiation}.
   
Many research has been conducted to simulate blind deployment that user and environmental conditions trigger.
Daysim tool for annual daylight availability and influence of automated lighting controls~\cite{reinhart2001validation}, daylight responsive dimmer for electric lighting~\cite{athienitis2002methodology}, automated blind control for user comfort~\cite{wienold2007dynamic,wienold2009dynamic,koo2010automated}, energy consumption triggered by luminaire dimming and occupancy sensors~\cite{roisin2008lighting}, and considerations of multiple variables are a few examples of studies looking at activation~\cite{daum2010identifying}.
   
A building's overall energy consumption will be impacted by occupant interactions and activities, but few models take these activities' effects on each parameter into account~\cite{hoes2009user}.
Interaction experiments on lighting simulation are concentrated on light switching and shade activation.
To forecast light switching, window opening, and blind shutting associated to the time of year, temperature, and number of occupants, algorithms based on real office studies were employed~\cite{reinhart2004lightswitch,herkel2008towards,lindelof2006field,bourgeois2006adding}.

\textbf{Designing new components:}
Even though predictive rendering adheres to physical rules, it cannot replicate every physically feasible aspect.
This is especially true for components like light concentrators that operate on principles not included in the geometrical optical model.
Additionally, it can be challenging to characterize some components due to the calculating techniques utilized by algorithms.
For instance, modeling complicated curved and specular surfaces using the Radiance model's original formulation is difficult or impossible due to backward raytracing and geometrical issues~\cite{ward1998rendering}.
Although large-diameter light pipes are still handled successfully~\cite{mohelnikova2008daylight}, thin light pipes and optic fiber are best described using the photon map approach~\cite{farrel2004lightpipe}.
It makes sense to assume that, as of this writing, additional initiatives are being taken to enhance the Radiance model and offer a trustworthy approach to emulating these components.
   
Specific components need to be tested in the lab using goniophotometers for the existing geometrical optical model to describe visual qualities adequately.
As in translucent panels~\cite{reinhart2006development} and many intricate fenestration systems, BRDF and BTDF functions are determined (CFS).

\textbf{Element prototyping:}
Several changes must be made to critical factors to find the ideal or nearly ideal element sizes.
Many researchers had to do simulations that involved trial and error to accomplish this (e.g.,~\cite{joarder2009simulation,li2005analysis,reinhart2005simulation,capeluto2003influence}.
Optimization software is the ideal method to utilize when there are too many test parameters.
However, the type of optimization algorithm, connectivity with optimizer software, and input received by the lighting simulation program must be taken into account.
Facades have been the subject of optimization studies because of their intricacy and impact on lighting and energy~\cite{wright2009geometric}.
They include ``random'' window placement for low energy consumption, intelligent facade optimization~\cite{park2003daylighting}, lighting controllers~\cite{guillemin2001innovative}, visual discomfort and solar penetration, external horizontal shading devices and their angles, and solar penetration~\cite{torres2007facade}.

\subsubsection{Input in lighting simulation}

Input will be outlined as the method by which data are extracted from reality and abstracted for processing by a simulation model.
The desired output quality and model behavior might be affected by input techniques.
Text files, command-line interfaces, translators from computer-aided drawing (CAD) tools, and graphical user interfaces (GUI) with or without their CAD system are some of the often used techniques.
Each input technique has benefits and drawbacks.
Despite using low computer resources, text files can be challenging to grasp.
GUIs require careful development because they range in complexity.
Software applications with their CAD systems can use DXF files from various CAD systems.
These GUI variations are ``plug-ins'' for well-known computer sketching programs~\cite{bleicher2009su2rad}.
While other techniques, such as GUI input, tend to assume variables for the user, input methods like text files and command prompt allow direct manipulation of model variables.
Results may be impacted by input methods' abstraction or simplification of reality.
For instance, curves are typically depicted using a lot of polygons.
Reflectance must be measured or weighted for exterior scene elements.
Sky models are used to describe how the length of the day changes, although there is a chance for a mistake, especially when estimating the brightness of a sunny sky~\cite{galasiu2002applicability}.

High dynamic range (HDR) photos can be utilized for scene analysis thanks to advances in digital photography.
Despite the need for specialized hardware and software, the generated photographs enable a pixel-by-pixel analysis of many illumination parameters~\cite{inanici2003transformation}.
Modern simulation models make it relatively simple to specify typical lighting, glazing, and fenestration systems.
Programs like WIN~\cite{ochoastate} or WIS~\cite{ochoa2010current} can create customized glazing combinations based on manufacturer databases for lighting and full-building simulations.
The accurate definition of optical characteristics, such as bidirectional transmittance distribution functions (BTDF), is necessary for other system types, such as complex fenestration systems (CFS)~\cite{reinhart2006development}.
These are identified using goniophotometer testing in a lab.
Translucent panels~\cite{reinhart2006development}, curved reflective blinds~\cite{andersen2005bi}, prismatic films~\cite{thanachareonkit2006comparing}, and laser-cut films are examples of elements that fit this description.
There is virtual goniophotometer software; however, it needs experimental validation~\cite{krishnaswamy2002virtual}.

\subsubsection{Output in lighting simulation}

Humans will interpret the output as the way the model's results are presented.
The output can be separated into two main categories: graphical representation and text-only (quantitative work) (qualitative output).
Each lighting simulation program offers a method for setting up output, such as choosing a format for text or graphics.
Photometric data from points on a previously established computation grid are typically contained in text-only files.
Results can be exported to word processors or spreadsheets.
These external programs modify data or display it (for example, in statistical analysis) (e.g., through diagrams).
Most often, the data contained in text-only files is strictly numerical and presented in tabular format.
Manipulating the simulation parameters can impact the accuracy of the numerical findings.
There are various uses for quality output.
A few instances include user preferences for illumination across a variety of ages~\cite{oi2005difference}, evaluation of user preferences for luminaire placement in offices~\cite{newsham2005lighting}, and simulation of complex visual stimuli for patients in the medical field~\cite{ruppertsberg2006rendering}.
Interactive images, rendering images, and images with data interpretation are three categories of quality output.

The output data provided by current techniques are diverse, but findings require a high level of knowledge to interpret~\cite{hong2000building}.
There are only so many tools for lighting simulation that give result interpretation or analysis~\cite{reinhart2006findings}.
Interpretation of the output should be tailored to the user type.
These come in two primary categories: One category consists of architects who don't have a background in lighting, but research how natural and artificial light affect their projects.
The second group includes lighting/daylighting researchers and physicists who favor validation and verification but are part-time designers~\cite{attia2009architect}.

\subsubsection{Applications of lighting simulation in architecture design process}

In this section we study the application of lighting simulation in architecture design process in order to provide new possibilities to improve the process of designing lighting system and also new features to maximize the quality of the users' experience.

\paragraph{Early stages of design}
Most computer tools available today are not suitable for use in the early phases of architectural design.
They require for a level of precision and detail that wasn't recognized at the time~\cite{aliakseyeu2006computer,sarawgi2004determining}.
Few made analyses or solution suggestions~\cite{reinhart2006findings}.
They are single-output based~\cite{hien2003computer} and targeted towards expert usage~\cite{petersen2010method}.
Taking too long to compute is a drawback~\cite{sarawgi2004determining}.
During the initial design stages, architects must interactively compare how their goals turned out.
    
Lightsolve~\cite{cutler2008interactive,andersen2008intuitive} was suggested for design exploration.
It provides direction on how particular cutting-edge fenestration solutions impact specific scenarios.
Their goal is to implement a goal-based design approach~\cite{lee2009goal}.
Sheng et al. presented a virtual heliodon as a ``sketcher'' that adjusts the intensity with altitude and acts like a natural sun~\cite{raskar1999spatially}.
Realistic illuminance forecasts were made in this manner.
A suggested setting for immersive VR is called SPOT~\cite{bund2005spot}.
Users may see the impact of altered building geometry on sunlight.
The facade combinations recommended by NewFacades~\cite{ochoa2009advice} show how they affect visual comfort and overall building energy.
Online overviews of how various lighting configurations seem under different lighting circumstances are offered by the Virtual Lighting Simulator and Daylight Variations Book~\cite{diepens2000daylight}.
    
Some models propose volumetric solutions that guarantee sun access to nearby structures.
They may be applied to urban planning and design's first stages.
Solar masks and programs like Sustarc~\cite{capeluto2001use}, Helios~\cite{seong2006helios}, and SunScapes~\cite{ratti2005sunscapes} allow users to compute the ``solar rights'' and ``solar envelopes'' concepts~\cite{alzoubi2010low}.
These techniques were first created for very bright climes, where a balance between summertime shade and wintertime access to solar radiation must be achieved.
    
Models may provide architectural concepts a tangible validity.
A range of maximum building heights for high rises is suggested by the tool by Ng, guaranteeing daylight access for lower levels~\cite{ng2005study}.
By figuring out the vertical daylight factor, it does this.
Capeluto et al. advocate angled urban layouts that also offer high density while ensuring sunlight access to all levels based on their solar envelopes tool~\cite{capeluto2005computer,capeluto2003climatic}.
According to shading~\cite{yezioro2006design}, the dimensions of urban squares or atriums should be chosen to maximize solar insolation and the placement of urban vegetation.
    
\paragraph{Design development(DD) stage}
After settling on fundamental concerns like massing, building placement, window size, and orientation, simulation models for lighting can be applied.
The following thoroughly explains the factors and procedures for effective daylight simulation~\cite{ochoastate}.
Tools created specifically for this purpose make it easier to explore artificial lighting configurations.
Solar shading studies may be completed quickly with CAD tools like AutoCAD and SketchUp.
SketchUp plug-ins, such as SunTools~\cite{ochoa2012state} offer more accurate daylight and electric lighting studies within CAD applications.
A building's unique components include glazing and redirection systems~\cite{page2007site}, shades and blinds~\cite{al2001computer}, and automated shading and lighting control modes to operate to their full potential, experts are required~\cite{kim2009manual,koo2010automated}.
    
\paragraph{Adaptation to building codes} 
Lighting simulation can be used to validate code compliance once a building project is finished and submitted for review by relevant authorities.
The simulation will become crucial as performance codes replace prescriptive ones~\cite{hien2003computer}.
Models of solar rights offer a framework for approving volume concepts.
Steward and Donn presented a technique to determine whether or not extensive simulations are required for compliance with interior daylight guidelines~\cite{stewart2007daylight}.
Sky view factor and view solid angles can be used to analyze window access to direct sun radiation~\cite{capeluto2003influence,souza2003sky}.
Compared to the traditional 60-degree obstruction restrictions, these methods successfully determine compliance.
    
\paragraph{Building commissioning and functioning}
Programs that simulate illumination are rarely used to simulate building operations.
As previously stated, it is challenging to portray human involvement and occupancy using various control components like blinds, moveable shades, sensors, and switches~\cite{lam2009occupancy,clear2006subject,inkarojrit2007multivariate}.
Doulos et al. provide a study and comparison of the most recent modeling methods for lighting control systems and daylight sensors~\cite{doulos2005critical}.
They discuss current switching patterns and their impact on energy use and evaluate various simulation techniques.
Daysim, one of these tools, analyzes different tactics employing control aspects.
Ehrlich et al. suggested radiation modeling of photosensors to forecast their performance and account for locational impacts on walls or ceilings~\cite{ehrlich2001simulating}.
Degelman proposed utilizing Monte Carlo probability to simulate motion sensors and anticipate light switching~\cite{degelman1999model}.
In whole-building simulation models that include both artificial and natural illumination, occupancy, switches, and sensors can be represented in varying degrees of abstraction.

\paragraph{Architecture specific lighting simulation}
Rendering an architecture scene is the most important part of the lighting simulation process since the visual presentation of the proposed environment will be rated by users and will affect their experience in the study.
This pressures the render engines to be optimized as possible with visual quality configurable during the experience.
The rendering takes the majority of update time since it will update by each change in real time so the choice of the most efficient tools to optimize the process.

For the creation of such experience, game engines are sophisticated, multifaceted tools.
They provide a setting for effective development, occasionally even without scripting skills.
The design of the user interface, rendering, physics, audio, animation, and other aspects of the game development process should all be covered by the game engine.

The game engines are in rivalry with one another and also comparing them is a complex undertaking because they are complicated tools.
Unity has been the most popular game engine in since 2016.
Unity is so simple to use that anyone can start producing with it; thus, it's likely to stay this way.
The Unreal Engine, which only has 17\% of Unity's users, is its main rival.
So, in this thesis, I'm contrasting those two engines.
The Cry Engine is also quite effective and used.
However, it is a fairly sophisticated program that is difficult for amateurs to use; therefore, studios are the primary users.

\textbf{Unity 3D Engine:}
The Unity game engine, developed by Unity Technologies Corporation, is currently at LTS version 2022.3.18f1 as of January 24, 2024.
Unity holds the distinction of being the most widely adopted game engine in contemporary usage, a position it has consistently maintained.
Its initial version was initially designed exclusively for OS X development, debuting at an Apple conference in 2005.
Over time, Unity has evolved and expanded its compatibility, becoming accessible on 27 distinct platforms, including extensive support for Virtual Reality (VR).

Unity's primary strength lies in its remarkable user-friendliness, which continues to be a pivotal feature of the engine.
The engine excels in accommodating rapid development, particularly within the realm of mobile platforms, as exemplified by the extensive reference~\cite{morse2021gaming}.
Unity projects and games tend to be relatively compact in scale, while the export process remains comparably straightforward.
Its component-based design philosophy simplifies the comprehension of its architecture.
For scripting, Unity employs C\#, an agile and efficient programming language that expedites the development process.

One of Unity's distinguishing attributes is its substantial user base, supplemented by vibrant online forums brimming with insightful solutions that aid in streamlining the debugging process.
Since its initial release, Unity has consistently faced challenges in the domain of graphics compared to rival game engines, notably Unreal and Cry Engine.
Although Unity introduced advanced features like the Unity Standard shader, physically based shading, and image-based lighting techniques in its Unity 5 version, it still grapples with certain shortcomings in comparison.

Regarding its lighting setup, Unity adopts a straightforward approach.
A solitary directional light source simulates the sun, casting soft shadows and imparting a subtle yellowish tint to the scene.
The sky plays a pivotal role in illuminating the landscape, with Unity employing a spherical High Dynamic Range (HDR) representation of the sky as ambient light.
An intriguing feature of Unity's lighting is its capacity to pre-calculate real-time global illumination, accomplished through the storage of a light map detailing the direction and intensity of indirect light.
The dynamic adjustment of light attributes, such as brightness, color, position, or direction, does not necessitate the time-consuming process of rebaking the lighting.
This real-time global illumination solution in Unity is aptly named ``Enlighten'' and is developed by the company Geometrics, offering significant advantages in lighting quality and efficiency~\cite{10108790}.

\textbf{Unreal Engine:}
Unreal Engine, a software framework developed by Epic Games, has gained prominence for its versatile capabilities and is currently at version 5.3 as of as of January 24, 2024.
Notably, it excels in facilitating the creation of immersive and realistic visual content, particularly in the realms of lifelike imagery, intricate vegetation, and complex terrain modeling.
Consequently, Unreal Engine is particularly well-suited for handling larger and more intricate projects that demand advanced graphical fidelity and computational resources.

While Unreal Engine provides support for deployment on iOS and Android platforms, it may not be the most suitable choice for the development of compact mobile games owing to its inherent complexity and resource-intensive nature.
To streamline the process of game development, Unreal Engine employs the Blueprint system, a visual scripting paradigm that enables developers to construct logic through the arrangement of interconnected blocks within graphical representations known as blueprints.
This approach eschews traditional script-based programming, offering an accessible and intuitive means of designing intricate logic.

One of Unreal Engine's standout features is its accessibility to the C++ source code, affording developers extensive control and customization opportunities over the entire system, harnessing the power of the C++ programming language.
However, it is noteworthy that mastering the foundational aspects of Unreal Engine can be challenging due to its inherent complexity.
Additionally, Unreal Engine boasts a robust and swift post-processing rendering technology that encompasses a rich array of features.

Furthermore, Unreal Engine provides a custom material editor, enabling users to craft intricate visual materials through a node-based system that simplifies shader programming using a visual interface. The engine's lighting capabilities are highly commendable, offering visually stunning lighting that can be either pre-rendered or delivered in real-time.
Unreal Engine's comprehensive toolset extends to visual debugging and optimization, offering valuable resources for enhancing project performance.

Nonetheless, it is worth noting that the Unreal Engine community is relatively smaller compared to some of its counterparts like Unity, and there may be gaps in available documentation and support.

Unreal Engine's enduring reputation in the field primarily hinges on its visual capabilities, which allow users to create intricate materials through a node-based shader programming approach with a visual interface.
The engine excels in rendering visually striking and dynamic lighting, offering the flexibility to choose between baked or real-time lighting solutions.
Notably, Unreal Engine distinguishes itself through its extensive suite of built-in post-processing effects, which significantly contribute to the overall visual quality.

In the context of architectural visualization, Unreal Engine frequently serves as the preferred platform for generating real-time walkthroughs, delivering exceptional fidelity and immersion in architectural representations.

The lighting arrangement is straightforward.
One directed light, the sun, which creates shadows in real-time, and one ambient light are used to illuminate the area and create an HDR sky.
Because we can configure the baking quality in the editor, the setup and baking are quicker than in Unity.
Still, each time we change, we have to recreate the lighting.
Unity doesn't need to regenerate lighting after making light modifications.
This method is also offered for Unreal, but it costs money.

The Unreal engine implements an optimization of shadows called Cascade Shadows.
According to the camera distance, it creates different shadow maps so that objects near the camera have sharper shadows than those farther away.
Additionally, there is a beautiful blend between the shadow maps, making the map change practically imperceptible.
This method is effective, especially for the more significant scenes~\cite{ha2022game}.

\textbf{Comparison:}
Table~\ref{tab:my-table} gives a brief comparison with a square table that lists the most routine duties engines have, together with a summary of the solutions each engine provides.
Both engines are strong and offer a complete set of resources for game development.
Which tool the developer prefers is usually a matter of personal preference.
There are several places, though, where one of the engines appears to offer a better or more comprehensive solution:

\begin{itemize}
\item
Unity supports many more target platforms, but Unreal also has the most important of them.
\item
Unreal's material editor allows the developer to design new shaders while considering materials.
Compared to the Unity Standard Shader, this strategy is significantly more potent.
\item
Unity supports pre-baked real-time global illumination, but Unreal doesn't provide this feature.
\item
Cascade particle system in Unreal is considerably better.
\item
Unreal has excellent built-in post-processing effects.
Similar effects are only possible with third-party plugins in Unity.
\end{itemize}

Considering this comparison, Unreal seems to be a more efficient simulation tool for the purpose of this review~\cite{sergeenko2022unity, christopoulou2017overview}.

\begin{table*}[ht!]
\caption{Comparison of Unity Game Engine and Unreal Engine.}
\label{tab:my-table}
\scriptsize
\centering
\resizebox{.85\textwidth}{!}{%
\renewcommand{\arraystretch}{1.5}
\begin{tabular}{|l|l|l|l|l|}
\cline{1-4}
\multicolumn{1}{|l}{} &  & \textbf{Unity} & \textbf{Unreal} \\ \hline
\multirow{4}{*}{Platform} & Desktop & Windows, macOS, Linux, WebGL & Windows, macOS, Linux, HTML 5 \\ \cline{2-4} 
 & Mobile & \begin{tabular}[c]{@{}l@{}}Windows phone, iOS, \\ Android, Blackberry 10, Tizen\end{tabular} & iOS, Android \\ \cline{2-4} 
 & Consoles & Xbox, Playstation, Nintendo & Xbox, Playstation, Nintendo \\ \cline{2-4} 
 & VR & \begin{tabular}[c]{@{}l@{}}SteamVR/HTC Vive, \\ Oculus Rift, Google VR/Daydream, \\ Samsung Gear VR, Hololens, Playstation VR\end{tabular} & \begin{tabular}[c]{@{}l@{}}SteamVR/HTC Vive, Oculus Rift, \\ Google VR/Daydream, \\ Samsung Gear VR, Playstation VR, OSVR\end{tabular} \\ \hline
 \multirow{4}{*}{Editor} & Tools & Unity Editor & Unreal Editor, VR Editor for HTC Vive \\ \cline{2-4} 
 & Features & Scene view and Game view as two windows  & Possess \& Eject, Simulate, Content browser \\ \cline{2-4} 
 & Scripting language & \begin{tabular}[c]{@{}l@{}}C\#, Unity Script,\\  Java Script, Python\end{tabular} & \begin{tabular}[c]{@{}l@{}}C++, Blueprint, \\ Visual Scripting\end{tabular} \\ \cline{2-4} 
 & Engine Code & Closed & C++ Source code available via Github \\ \hline
\multirow{7}{*}{Rendering} & Shading technique & Deferred Shading & Deffered Shading and Forward Shading for VR \\ \cline{2-4} 
 & Materials & \begin{tabular}[c]{@{}l@{}}Physically based, Standard Shader, \\ Tesselation Shader, Mobile Shader\end{tabular} & \begin{tabular}[c]{@{}l@{}}Physically based, Blueprint, Material editor, \\ Tesselation, Layred materials, Material instances, \\ Lit translucency, Subsurface shading model\end{tabular} \\ \cline{2-4} 
 & Lighting & \begin{tabular}[c]{@{}l@{}}Directional light, Point light, \\ Spot light, Area light\end{tabular} & \begin{tabular}[c]{@{}l@{}}Directional light, Point light, \\ Spot light, Sky light, \\ IES light profiles\end{tabular} \\ \cline{2-4} 
 & Shadows & Realtime, Hard/Soft shadows & \begin{tabular}[c]{@{}l@{}}Real time, Hard/Soft shadows, \\ Cascade shadow, Distance field shadows\end{tabular} \\ \cline{2-4} 
 & Global Illumination & \begin{tabular}[c]{@{}l@{}}Image-based GI, Precomputed realtime GI, \\ Baked GI\end{tabular} & Image-based GI, Baked GI \\ \cline{2-4} 
 & Reflection/Refraction & Reflection probes & \begin{tabular}[c]{@{}l@{}}Reflection probes, Screen space, \\ Reflections, Refraction\end{tabular} \\ \cline{2-4} 
 & Particles & Sample curves based particle system & Cascade particle system with GPU particles \\
 \cline{1-4}
\end{tabular}%
}
\end{table*}

\begin{table*}[htbp]
\caption{Simple simulation tools for artificial lighting technology in building applications and associated parameters --- Part 1 (multiple citations).}
\label{tab:simpleSim}
\scriptsize
\centering
\resizebox{0.85\paperwidth}{!}{%
\renewcommand{\arraystretch}{1.5}
\begin{adjustbox}{angle=0}
\begin{tabular}{ll|l|llll}
\cline{3-7}
\multicolumn{2}{l|}{} & \textbf{Paper \#} & \multicolumn{1}{l|}{\textbf{Tool}} & \multicolumn{1}{l|}{\textbf{Features}} & \multicolumn{1}{l|}{Light Source} & \multicolumn{1}{l|}{Building Type} \\ \cline{3-7}
\multicolumn{2}{l|}{} & \multirow{6}{*}{20} & \multicolumn{1}{l|}{\multirow{6}{*}{MATLAB~\cite{matlab_shaikh2018intelligent,matlab_shaikh2013indoor,matlab_alcala2003fuzzy,matlab_ali2013energy,matlab_guillemin2002energy,matlab_kristl2008fuzzy,matlab_lah2005fuzzy,matlab_lah2006daylight,matlab_park2003daylighting,matlab_penacchio2015visual,matlab_rodriguez2015fuzzy,matlab_rutishauser2004control,matlab_shaikh2013building,matlab_shaikh2013robust,matlab_shaikh2014stochastic,matlab_shaikh2016intelligent,matlab_wang2012intelligent,matlab_wang2012multi,matlab_wang2012occupancy,366462933}}} & \multicolumn{1}{l|}{Energy efficiency} & \multicolumn{1}{l|}{\multirow{6}{*}{Natural and Artificial}} & \multicolumn{1}{l|}{\multirow{6}{*}{\begin{tabular}[c]{@{}l@{}}Commercial\\      Industrial\\      Office\\      Residential\end{tabular}}} \\
\multicolumn{2}{l|}{} &  & \multicolumn{1}{l|}{} & \multicolumn{1}{l|}{Illuminance} & \multicolumn{1}{l|}{} & \multicolumn{1}{l|}{} \\
\multicolumn{2}{l|}{} &  & \multicolumn{1}{l|}{} & \multicolumn{1}{l|}{Specific   utility function} & \multicolumn{1}{l|}{} & \multicolumn{1}{l|}{} \\
\multicolumn{2}{l|}{} &  & \multicolumn{1}{l|}{} & \multicolumn{1}{l|}{User comfort   parameters} & \multicolumn{1}{l|}{} & \multicolumn{1}{l|}{} \\
\multicolumn{2}{l|}{} &  & \multicolumn{1}{l|}{} & \multicolumn{1}{l|}{User   expectations and demands} & \multicolumn{1}{l|}{} & \multicolumn{1}{l|}{} \\
\multicolumn{2}{l|}{} &  & \multicolumn{1}{l|}{} & \multicolumn{1}{l|}{User   preference} & \multicolumn{1}{l|}{} & \multicolumn{1}{l|}{} \\ \cline{3-7} 
\multicolumn{2}{l|}{} & \multirow{6}{*}{8} & \multicolumn{1}{l|}{\multirow{6}{*}{MATLAB/Simulink~\cite{matsim_wang2013illumination,matsim_asare2012integrated,matsim_wang2011intelligent,matsim_wang2010multi,matsim_kolokotsa2003comparison,matsim_kolokotsa2002genetic,matsim_guillemin2001innovative,matsim_kolokotsa2001advanced}}} & \multicolumn{1}{l|}{Energy efficiency} & \multicolumn{1}{l|}{\multirow{6}{*}{Natural   and Artificial}} & \multicolumn{1}{l|}{\multirow{6}{*}{\begin{tabular}[c]{@{}l@{}}Commercial\\      Industrial\\      Office\\      Residential\end{tabular}}} \\
\multicolumn{2}{l|}{} &  & \multicolumn{1}{l|}{} & \multicolumn{1}{l|}{Illuminance} & \multicolumn{1}{l|}{} & \multicolumn{1}{l|}{} \\
\multicolumn{2}{l|}{} &  & \multicolumn{1}{l|}{} & \multicolumn{1}{l|}{Occupant   patterns} & \multicolumn{1}{l|}{} & \multicolumn{1}{l|}{} \\
\multicolumn{2}{l|}{} &  & \multicolumn{1}{l|}{} & \multicolumn{1}{l|}{User behavior} & \multicolumn{1}{l|}{} & \multicolumn{1}{l|}{} \\
\multicolumn{2}{l|}{} &  & \multicolumn{1}{l|}{} & \multicolumn{1}{l|}{User   occupancy and preference request} & \multicolumn{1}{l|}{} & \multicolumn{1}{l|}{} \\
\multicolumn{2}{l|}{} &  & \multicolumn{1}{l|}{} & \multicolumn{1}{l|}{Visual   comfort} & \multicolumn{1}{l|}{} & \multicolumn{1}{l|}{} \\ \cline{3-7} 
\multicolumn{2}{l|}{} & \multirow{5}{*}{5} & \multicolumn{1}{l|}{\multirow{5}{*}{EenergyPlus~\cite{eneryp_delgarm2016novel,eneryp_dong2009sensor,eneryp_pisello2012building,eneryp_liu2012simulation,eneryp_costanzo2016thermal}}} & \multicolumn{1}{l|}{Energy efficiency} & \multicolumn{1}{l|}{\multirow{5}{*}{Natural}} & \multicolumn{1}{l|}{\multirow{5}{*}{\begin{tabular}[c]{@{}l@{}}Commercial\\      Industrial\\      Office\end{tabular}}} \\
\multicolumn{2}{l|}{} &  & \multicolumn{1}{l|}{} & \multicolumn{1}{l|}{Illuminance} & \multicolumn{1}{l|}{} & \multicolumn{1}{l|}{} \\
\multicolumn{2}{l|}{} &  & \multicolumn{1}{l|}{} & \multicolumn{1}{l|}{Occupants'   attitude/satisfaction level} & \multicolumn{1}{l|}{} & \multicolumn{1}{l|}{} \\
\multicolumn{2}{l|}{} &  & \multicolumn{1}{l|}{} & \multicolumn{1}{l|}{User-specified   time steps} & \multicolumn{1}{l|}{} & \multicolumn{1}{l|}{} \\
\multicolumn{2}{l|}{} &  & \multicolumn{1}{l|}{} & \multicolumn{1}{l|}{Visual,   thermal and cooling comfort} & \multicolumn{1}{l|}{} & \multicolumn{1}{l|}{} \\ \cline{3-7} 
\multicolumn{2}{l|}{} & \multirow{4}{*}{4} & \multicolumn{1}{l|}{\multirow{4}{*}{MATLAB/GUI~\cite{matgui_yang2011gui,matgui_yang2012multi,matgui_yang2013multi,matgui_baloch2016graphical}}} & \multicolumn{1}{l|}{Energy efficiency} & \multicolumn{1}{l|}{\multirow{4}{*}{Natural   and Artificial}} & \multicolumn{1}{l|}{\multirow{4}{*}{\begin{tabular}[c]{@{}l@{}}Commercial\\      Industrial\\      Office\\      Residential\end{tabular}}} \\
\multicolumn{2}{l|}{} &  & \multicolumn{1}{l|}{} & \multicolumn{1}{l|}{Illuminance} & \multicolumn{1}{l|}{} & \multicolumn{1}{l|}{} \\
\multicolumn{2}{l|}{} &  & \multicolumn{1}{l|}{} & \multicolumn{1}{l|}{Occupants preference} & \multicolumn{1}{l|}{} & \multicolumn{1}{l|}{} \\
\multicolumn{2}{l|}{} &  & \multicolumn{1}{l|}{} & \multicolumn{1}{l|}{Visual comfort} & \multicolumn{1}{l|}{} & \multicolumn{1}{l|}{} \\ \cline{3-7} 
\multicolumn{2}{l|}{} & \multirow{3}{*}{2} & \multicolumn{1}{l|}{\multirow{3}{*}{DOE~\cite{doe_siddharth2011automatic,doe_ghisi2001optimising}}} & \multicolumn{1}{l|}{Energy (Window area)} & \multicolumn{1}{l|}{\multirow{3}{*}{Artificial}} & \multicolumn{1}{l|}{\multirow{3}{*}{Office}} \\
\multicolumn{2}{l|}{} &  & \multicolumn{1}{l|}{} & \multicolumn{1}{l|}{Illuminance} & \multicolumn{1}{l|}{} & \multicolumn{1}{l|}{} \\
\multicolumn{2}{l|}{} &  & \multicolumn{1}{l|}{} & \multicolumn{1}{l|}{User defined/Input parameters} & \multicolumn{1}{l|}{} & \multicolumn{1}{l|}{} \\ \cline{3-7} 
\multicolumn{2}{l|}{} & \multirow{4}{*}{2} & \multicolumn{1}{l|}{\multirow{4}{*}{JAVA~\cite{java_paiz2008hardware,java_wen2008wireless}}} & \multicolumn{1}{l|}{Energy efficiency} & \multicolumn{1}{l|}{\multirow{4}{*}{NA}} & \multicolumn{1}{l|}{\multirow{4}{*}{\begin{tabular}[c]{@{}l@{}}Commercial\\      Office\\      Residential\end{tabular}}} \\ 
\multicolumn{2}{l|}{} &  & \multicolumn{1}{l|}{} & \multicolumn{1}{l|}{Illuminance} & \multicolumn{1}{l|}{} & \multicolumn{1}{l|}{} \\
\multicolumn{2}{l|}{} &  & \multicolumn{1}{l|}{} & \multicolumn{1}{l|}{User satisfaction / preference} & \multicolumn{1}{l|}{} & \multicolumn{1}{l|}{} \\
\multicolumn{2}{l|}{} &  & \multicolumn{1}{l|}{} & \multicolumn{1}{l|}{Visual comfort} & \multicolumn{1}{l|}{} & \multicolumn{1}{l|}{} \\ \cline{3-7} 

\multicolumn{2}{l|}{} & \multirow{2}{*}{2} & \multicolumn{1}{l|}{\multirow{2}{*}{Daysim~\cite{daysim_acosta2017analysis,daysim_acosta2016window}}} & \multicolumn{1}{l|}{\multirow{2}{*}{Energy and ligthing}} & \multicolumn{1}{l|}{\multirow{2}{*}{Natural and Daylight}} & \multicolumn{1}{l|}{\multirow{2}{*}{\begin{tabular}[c]{@{}l@{}}Office\\ Residential\end{tabular}}} \\
\multicolumn{2}{l|}{} &  & \multicolumn{1}{l|}{} & \multicolumn{1}{l|}{} & \multicolumn{1}{l|}{} & \multicolumn{1}{l|}{} \\ \cline{3-7}  
\end{tabular}%
\end{adjustbox}
}
\end{table*}

\begin{table*}[htbp]
\caption{Simple simulation tools for artificial lighting technology in building applications and associated parameters --- Part 2 (single citation).}
\label{tab:simpleSim2}
\scriptsize
\centering
\resizebox{.85\paperwidth}{!}{%
\renewcommand{\arraystretch}{1.5}
\begin{adjustbox}{angle=0}
\begin{tabular}{ll|l|llll}
\cline{3-7}
\multicolumn{2}{l|}{} & Paper \# & \multicolumn{1}{l|}{Tool} & \multicolumn{1}{l|}{Features} & \multicolumn{1}{l|}{Light Source} & \multicolumn{1}{l|}{Building Type} \\ \cline{3-7}
\multicolumn{2}{l|}{} & \multirow{6}{*}{1} & \multicolumn{1}{l|}{\multirow{4}{*}{CLIPS~\cite{clips_doukas2007intelligent}}} & \multicolumn{1}{l|}{Energy efficiency} & \multicolumn{1}{l|}{\multirow{4}{*}{NA}} & \multicolumn{1}{l|}{\multirow{4}{*}{\begin{tabular}[c]{@{}l@{}}Commercial\\      Industrial\\      Office\\      Residential\end{tabular}}} \\
\multicolumn{2}{l|}{} &  & \multicolumn{1}{l|}{} & \multicolumn{1}{l|}{Illuminance} & \multicolumn{1}{l|}{} & \multicolumn{1}{l|}{} \\
\multicolumn{2}{l|}{} &  & \multicolumn{1}{l|}{} & \multicolumn{1}{l|}{User inputs / Requirements} & \multicolumn{1}{l|}{} & \multicolumn{1}{l|}{} \\
\multicolumn{2}{l|}{} &  & \multicolumn{1}{l|}{} & \multicolumn{1}{l|}{} & \multicolumn{1}{l|}{} & \multicolumn{1}{l|}{} \\ \cline{3-7} 
\multicolumn{2}{l|}{} & \multirow{5}{*}{1} & \multicolumn{1}{l|}{\multirow{5}{*}{BuildOpt~\cite{buildopt_wetter2005building}}} & \multicolumn{1}{l|}{Energy efficiency} & \multicolumn{1}{l|}{\multirow{5}{*}{NA}} & \multicolumn{1}{l|}{\multirow{5}{*}{Office}} \\
\multicolumn{2}{l|}{} &  & \multicolumn{1}{l|}{} & \multicolumn{1}{l|}{Shading} & \multicolumn{1}{l|}{} & \multicolumn{1}{l|}{} \\
\multicolumn{2}{l|}{} &  & \multicolumn{1}{l|}{} & \multicolumn{1}{l|}{User-specified limits} & \multicolumn{1}{l|}{} & \multicolumn{1}{l|}{} \\
\multicolumn{2}{l|}{} &  & \multicolumn{1}{l|}{} & \multicolumn{1}{l|}{Visual comfort} & \multicolumn{1}{l|}{} & \multicolumn{1}{l|}{} \\
\multicolumn{2}{l|}{} &  & \multicolumn{1}{l|}{} & \multicolumn{1}{l|}{Windows dimension} & \multicolumn{1}{l|}{} & \multicolumn{1}{l|}{} \\ \cline{3-7} 
\multicolumn{2}{l|}{} & \multirow{2}{*}{1} & \multicolumn{1}{l|}{\multirow{2}{*}{LabVIEW~\cite{labview_yoon2011line}}} & \multicolumn{1}{l|}{Energy efficiency} & \multicolumn{1}{l|}{\multirow{2}{*}{NA}} & \multicolumn{1}{l|}{\multirow{2}{*}{Office}} \\
\multicolumn{2}{l|}{} &  & \multicolumn{1}{l|}{} & \multicolumn{1}{l|}{Illuminance} & \multicolumn{1}{l|}{} & \multicolumn{1}{l|}{} \\ \cline{3-7} 
\multicolumn{2}{l|}{} & \multirow{3}{*}{1} & \multicolumn{1}{l|}{\multirow{3}{*}{LightWise~\cite{lightwise_delaney2009evaluation}}} & \multicolumn{1}{l|}{Consumer preference} & \multicolumn{1}{l|}{\multirow{3}{*}{NA}} & \multicolumn{1}{l|}{\multirow{3}{*}{Office}} \\
\multicolumn{2}{l|}{} &  & \multicolumn{1}{l|}{} & \multicolumn{1}{l|}{Energy efficiency} & \multicolumn{1}{l|}{} & \multicolumn{1}{l|}{} \\
\multicolumn{2}{l|}{} &  & \multicolumn{1}{l|}{} & \multicolumn{1}{l|}{Illuminance} & \multicolumn{1}{l|}{} & \multicolumn{1}{l|}{} \\ \cline{3-7} 
\multicolumn{2}{l|}{} & \multirow{3}{*}{1} & \multicolumn{1}{l|}{\multirow{3}{*}{Prototype Software Tool~\cite{protSoft_marinakis2013integrated}}} & \multicolumn{1}{l|}{Energy efficiency} & \multicolumn{1}{l|}{\multirow{3}{*}{Artificial}} & \multicolumn{1}{l|}{\multirow{3}{*}{Commercial}} \\
\multicolumn{2}{l|}{} &  & \multicolumn{1}{l|}{} & \multicolumn{1}{l|}{Illuminance} & \multicolumn{1}{l|}{} & \multicolumn{1}{l|}{} \\
\multicolumn{2}{l|}{} &  & \multicolumn{1}{l|}{} & \multicolumn{1}{l|}{User requirements and profile} & \multicolumn{1}{l|}{} & \multicolumn{1}{l|}{} \\ \cline{3-7} 
\multicolumn{2}{l|}{} & \multirow{3}{*}{1} & \multicolumn{1}{l|}{\multirow{3}{*}{Proteus~\cite{proteus_mishra2017wpan}}} & \multicolumn{1}{l|}{Energy efficiency} & \multicolumn{1}{l|}{\multirow{3}{*}{Artificial}} & \multicolumn{1}{l|}{\multirow{3}{*}{\begin{tabular}[c]{@{}l@{}}Commercial\\      Office\\      Residential\end{tabular}}} \\
\multicolumn{2}{l|}{} &  & \multicolumn{1}{l|}{} & \multicolumn{1}{l|}{Visual and other comforts} & \multicolumn{1}{l|}{} & \multicolumn{1}{l|}{} \\
\multicolumn{2}{l|}{} &  & \multicolumn{1}{l|}{} & \multicolumn{1}{l|}{} & \multicolumn{1}{l|}{} & \multicolumn{1}{l|}{} \\ \cline{3-7} 
\multicolumn{2}{l|}{} & 1 & \multicolumn{1}{l|}{Hemera~\cite{hemera_malet2017performance}} & \multicolumn{1}{l|}{Visual comfort} & \multicolumn{1}{l|}{Natural} & \multicolumn{1}{l|}{\begin{tabular}[c]{@{}l@{}}Office\\      Residential\end{tabular}} \\ \cline{3-7} 
\multicolumn{2}{l|}{} & \multirow{3}{*}{1} & \multicolumn{1}{l|}{\multirow{3}{*}{SEMPER~\cite{semper_mahdavi2003enclosure}}} & \multicolumn{1}{l|}{Energy efficiency} & \multicolumn{1}{l|}{\multirow{3}{*}{NA}} & \multicolumn{1}{l|}{\multirow{3}{*}{\begin{tabular}[c]{@{}l@{}}Commercial\\      Office\end{tabular}}} \\
\multicolumn{2}{l|}{} &  & \multicolumn{1}{l|}{} & \multicolumn{1}{l|}{Illuminance} & \multicolumn{1}{l|}{} & \multicolumn{1}{l|}{} \\
\multicolumn{2}{l|}{} &  & \multicolumn{1}{l|}{} & \multicolumn{1}{l|}{User input set values} & \multicolumn{1}{l|}{} & \multicolumn{1}{l|}{} \\ \cline{3-7} 
\multicolumn{2}{l|}{} & \multirow{3}{*}{1} & \multicolumn{1}{l|}{\multirow{3}{*}{Agent-Based Model Simulation~\cite{agent_erickson2009energy}}} & \multicolumn{1}{l|}{Energy efficiency} & \multicolumn{1}{l|}{\multirow{3}{*}{NA}} & \multicolumn{1}{l|}{\multirow{3}{*}{Office}} \\
\multicolumn{2}{l|}{} &  & \multicolumn{1}{l|}{} & \multicolumn{1}{l|}{Illuminance} & \multicolumn{1}{l|}{} & \multicolumn{1}{l|}{} \\
\multicolumn{2}{l|}{} &  & \multicolumn{1}{l|}{} & \multicolumn{1}{l|}{Occupancy sensor camera} & \multicolumn{1}{l|}{} & \multicolumn{1}{l|}{} \\ \cline{3-7} 
\multicolumn{2}{l|}{} & 1 & \multicolumn{1}{l|}{Integrated   Environmental Solution-Virtual Environment (IES-VE)~\cite{ies_lim2016dynamic}} & \multicolumn{1}{l|}{Visual comfort} & \multicolumn{1}{l|}{Natural} & \multicolumn{1}{l|}{Office} \\ \cline{3-7} 
\multicolumn{2}{l|}{} & \multirow{3}{*}{1} & \multicolumn{1}{l|}{\multirow{3}{*}{DIALux~\cite{dialux_soori2013lighting}}} & \multicolumn{1}{l|}{Energy saving} & \multicolumn{1}{l|}{\multirow{3}{*}{Natural, Artificial, and Daylight}} & \multicolumn{1}{l|}{\multirow{3}{*}{Office}} \\
\multicolumn{2}{l|}{} &  & \multicolumn{1}{l|}{} & \multicolumn{1}{l|}{Luminaire arrangments} & \multicolumn{1}{l|}{} & \multicolumn{1}{l|}{} \\
\multicolumn{2}{l|}{} &  & \multicolumn{1}{l|}{} & \multicolumn{1}{l|}{Occupants' comfort} & 
\multicolumn{1}{l|}{} & \multicolumn{1}{l|}{} \\ \cline{3-7}
\end{tabular}%
\end{adjustbox}
}
\end{table*}

\begin{table*}[htpb]
\caption{Integrated simulation tools for artificial lighting technology in building applications and associated parameters.}
\label{tab:integratedSim}
\scriptsize
\centering
\resizebox{.85\paperwidth}{!}{%
\renewcommand{\arraystretch}{1.5}
\begin{adjustbox}{angle=0}
\begin{tabular}{ll|l|llll}
\cline{3-7}
\multicolumn{2}{l|}{} & Paper \# & \multicolumn{1}{l|}{Tool} & \multicolumn{1}{l|}{Features} & \multicolumn{1}{l|}{Light Source} & \multicolumn{1}{l|}{Building Type} \\ \cline{3-7}
\multicolumn{2}{l|}{} & \multirow{3}{*}{4} & \multicolumn{1}{l|}{\multirow{3}{*}{EnergyPlus + Radiance~\cite{energy_rad_bustamante2017integrated,energy_rad_vera2017optimization,energy_rad_vera2016flexible,energy_rad_huang2015energy}}} & \multicolumn{1}{l|}{Energy and lighting} & \multicolumn{1}{l|}{\multirow{3}{*}{Natural,   Artificial, and Daylight}} & \multicolumn{1}{l|}{\multirow{3}{*}{\begin{tabular}[c]{@{}l@{}}Commercial\\      Office\end{tabular}}} \\
\multicolumn{2}{l|}{} &  & \multicolumn{1}{l|}{} & \multicolumn{1}{l|}{Thermal comfort} & \multicolumn{1}{l|}{} & \multicolumn{1}{l|}{} \\
\multicolumn{2}{l|}{} &  & \multicolumn{1}{l|}{} & \multicolumn{1}{l|}{Visual comfort/discomfort} & \multicolumn{1}{l|}{} & \multicolumn{1}{l|}{} \\ \cline{3-7} 
\multicolumn{2}{l|}{} & \multirow{2}{*}{2} & \multicolumn{1}{l|}{\multirow{2}{*}{EnergyPlus + MATLAB~\cite{energy_mat_kim2016simulation,energy_mat_ji2012energy}}} & \multicolumn{1}{l|}{Applied energy} & \multicolumn{1}{l|}{\multirow{2}{*}{Artificial}} & \multicolumn{1}{l|}{\multirow{2}{*}{Office}} \\
\multicolumn{2}{l|}{} &  & \multicolumn{1}{l|}{} & \multicolumn{1}{l|}{Energy saving in lighting} & \multicolumn{1}{l|}{} & \multicolumn{1}{l|}{} \\ \cline{3-7} 
\multicolumn{2}{l|}{} & 1 & \multicolumn{1}{l|}{MATLAB + LabVIEW~\cite{mat_lab_xiong2016model}} & \multicolumn{1}{l|}{Glare constraints} & \multicolumn{1}{l|}{Natural} & \multicolumn{1}{l|}{Office} \\ \cline{3-7} 
\multicolumn{2}{l|}{} & \multirow{2}{*}{1} & \multicolumn{1}{l|}{\multirow{2}{*}{MATLAB + EnergyPlus + BCVTB~\cite{mat_ene_bc_shen2014energy}}} & \multicolumn{1}{l|}{Energy efficiency} & \multicolumn{1}{l|}{\multirow{2}{*}{NA}} & \multicolumn{1}{l|}{\multirow{2}{*}{Office}} \\
\multicolumn{2}{l|}{} &  & \multicolumn{1}{l|}{} & \multicolumn{1}{l|}{Visual and other comfort   parameters} & \multicolumn{1}{l|}{} & \multicolumn{1}{l|}{} \\ \cline{3-7} 
\multicolumn{2}{l|}{} & \multirow{4}{*}{1} & \multicolumn{1}{l|}{\multirow{4}{*}{TRNSYS + MATLAB~\cite{dounis2007intelligent}}} & \multicolumn{1}{l|}{Illuminance} & \multicolumn{1}{l|}{\multirow{4}{*}{Artificial}} & \multicolumn{1}{l|}{\multirow{4}{*}{\begin{tabular}[c]{@{}l@{}}Commercial\\      Industrial\\      Office\\      Residential\end{tabular}}} \\
\multicolumn{2}{l|}{} &  & \multicolumn{1}{l|}{} & \multicolumn{1}{l|}{User preference and dependency} & \multicolumn{1}{l|}{} & \multicolumn{1}{l|}{} \\
\multicolumn{2}{l|}{} &  & \multicolumn{1}{l|}{} & \multicolumn{1}{l|}{} & \multicolumn{1}{l|}{} & \multicolumn{1}{l|}{} \\
\multicolumn{2}{l|}{} &  & \multicolumn{1}{l|}{} & \multicolumn{1}{l|}{} & \multicolumn{1}{l|}{} & \multicolumn{1}{l|}{} \\ \cline{3-7} 
\multicolumn{2}{l|}{} & \multirow{2}{*}{1} & \multicolumn{1}{l|}{\multirow{2}{*}{TRNSYS + INSEL~\cite{trn_ins_colmenar2013solutions}}} & \multicolumn{1}{l|}{Energy efficiency} & \multicolumn{1}{l|}{\multirow{2}{*}{Artificial}} & \multicolumn{1}{l|}{\multirow{2}{*}{Office}} \\
\multicolumn{2}{l|}{} &  & \multicolumn{1}{l|}{} & \multicolumn{1}{l|}{Illuminance} & \multicolumn{1}{l|}{} & \multicolumn{1}{l|}{} \\ \cline{3-7} 
\multicolumn{2}{l|}{} & \multirow{3}{*}{1} & \multicolumn{1}{l|}{\multirow{3}{*}{DRIVE + WSE + OSGI~\cite{dr_wse_osgi_chen2009design}}} & \multicolumn{1}{l|}{Energy efficiency} & \multicolumn{1}{l|}{\multirow{3}{*}{NA}} & \multicolumn{1}{l|}{\multirow{3}{*}{Office}} \\
\multicolumn{2}{l|}{} &  & \multicolumn{1}{l|}{} & \multicolumn{1}{l|}{Illuminance} & \multicolumn{1}{l|}{} & \multicolumn{1}{l|}{} \\
\multicolumn{2}{l|}{} &  & \multicolumn{1}{l|}{} & \multicolumn{1}{l|}{Occupants awareness} & \multicolumn{1}{l|}{} & \multicolumn{1}{l|}{} \\ \cline{3-7}
\end{tabular}%
\end{adjustbox}
}
\end{table*}

\subsection{RQ2: How can XR technology be used for SBE lighting system modeling?}

As underscored in in the background section, extended reality (XR) which is an all-encompassing term that comprises virtual reality (VR), augmented reality (VR), and mixed reality (MR), encompasses an assortment of cutting-edge technologies that merge the physical and digital worlds, providing users with immersive and interactive experiences.
XR can be visualized through portable devices such as smartphones and tablets, or through head mounted displays (HMDs)~\cite{casini_extended_2022}.

In this section, we explore each immersive technology in detail, emphasizing the pertinent research findings for each one, and offering illustrative examples of their applications in the field of smart lighting design.
VR, a computer-generated environment, offers a simulated world that is fully immersive, enclosing the user within a headset that obstructs the actual environment.
Through handheld controllers, users can manipulate objects and interact within the virtual world.
AR, on the other hand, superimposes digital content onto the user's real-world environment, often facilitated through a smartphone or tablet device.
AR utilizes the device's camera to capture the surroundings and integrate digital imagery, videos, or 3D models into the live stream.
MR is an integration of both virtual and augmented reality, allowing digital objects to blend seamlessly into the real world. MR headsets have cameras and sensors that track the user's surroundings, enabling digital objects to be placed and interact with the real world.

Lighting is a critical element in creating an immersive and realistic XR experience.
Light affects how we perceive objects, spaces, and materials.
It also helps create a sense of depth, texture, and atmosphere.
Without proper lighting, XR experiences can appear flat, lifeless, and unconvincing.
In XR applications, lighting can be used to convey mood, emotion, and tone, helping to create an emotional connection between the user and the virtual environment (VE).
For instance, in a VR horror game, dimly lit corridors with flickering lights can create an eerie and ominous atmosphere, increasing the user's fear and tension.
Similarly, in an AR museum exhibit, soft and warm lighting can create a calming and inviting ambiance, enhancing the user's ability to appreciate the art and artifacts on display.
In MR training simulations, lighting can affect the user's sense of presence and immersion in the scenario.
For instance, realistic lighting conditions in a virtual construction site can help trainees better understand the lighting and visibility challenges that they might face in the real world.

Integrating smart lighting design into XR environments is a promising approach to enhancing the user's experience by enabling more precise control over lighting conditions.
Advanced smart lighting systems equipped with sensors and algorithms can dynamically adjust lighting levels, color temperature, and directionality in real-time based on the user's position and movements.
This provides a more immersive and responsive experience, as lighting conditions can be optimized to create an environment that more accurately simulates real-world scenarios.
Moreover, smart lighting design can enhance the overall aesthetic of an XR environment by creating stunning visual effects through creative use of color, brightness, and contrast.
Additionally, energy-efficient lighting systems can help reduce the carbon footprint of XR environments, contributing to sustainability efforts.
By incorporating smart lighting design into XR environments, designers and developers can create more engaging and immersive experiences that provide users with a heightened sense of presence and realism.

In the realm of XR lighting, there are three primary types of lighting used in XR: ambient lighting, directional lighting, and point lighting.
Ambient lighting is the overall lighting in the environment, and it sets the tone for the scene.
This type of lighting is usually used to create a sense of space and mood.
In XR, ambient lighting can be used to simulate natural lighting conditions, such as daylight, dawn, dusk, and nighttime.
Directional lighting is used to create shadows and highlights on objects in the environment.
This type of lighting can be used to simulate the sun's position in the sky, as well as to create dramatic effects, such as silhouettes.
Point lighting is used to highlight specific objects or areas in the environment.
This type of lighting is often used to draw the user's attention to a particular object or feature in the environment.

Beyond these primary lighting methods, two additional techniques, linear lighting and surface lighting, contribute to crafting realistic 3D graphics in VE.
Linear lighting, also referred to as physically based rendering (PBR), replicates real-world lighting interactions by considering material properties such as reflectivity and roughness, yielding more genuine lighting effects.
Meanwhile, surface lighting techniques encompass various approaches, including masking-shadowing functions, subsurface scattering, and efficient irradiance environment mapping.

Ambient occlusion is one such surface lighting technique, mimicking soft shadows from nearby objects and adding depth and realism to XR scenes.
Shadow mapping ensures the precise rendering of object shadows, enhancing spatial perception by considering light occlusion.
High dynamic range (HDR) rendering broadens the display's brightness range, enabling the representation of bright highlights and intricate details in the VE, boosting overall visual fidelity.
Finally, environment mapping reflects the surrounding environment onto object surfaces, adding to the XR experience's realism and utility, particularly for reflective surfaces like mirrors or metallic objects.
These techniques collectively enrich the VE, offering a more immersive and visually appealing experience.

Researchers are actively exploring XR and global illumination approaches to expand and enhance options for studying interior lighting situations.
Several studies have been conducted to develop a sample VE that improves and evaluates lighting and design training.
These studies aim to offer a promising solution for increasing experimentation in a crowded experience test-bed and reducing the time required for user comprehension and understanding.
By utilizing VR and global illumination techniques, researchers can create immersive environments that accurately simulate real-world lighting conditions and enable the evaluation of various lighting scenarios.
Such approaches provide a valuable tool for designers, architects, and other professionals to test and refine their ideas and concepts, ultimately leading to more efficient and effective lighting design solutions.

The investigation of XR technologies in collaborative urban design and their potential to meet people's lighting needs is a new and emerging field of research.
As XR technologies become more advanced, there is an increasing need to understand how they can be utilized to create more engaging and immersive urban environments that meet the diverse needs of communities.
Research in this area aims to explore the use of VR and AR to simulate different lighting scenarios and provide designers, planners, and communities with a more comprehensive understanding of how lighting can affect the urban environment.
Lighting systems and its required parameters that may characterize the quantity, and quality of light are necessary in a VE~\cite{bellazzi_virtual_2022}.

Wittkamper et al.~\cite{970535} indicated lighting is a crucial design element that may be utilized to set the mood and ambiance of a space, highlight particular points of interest, or visually mute distracting elements.
Since lighting design is one of the XR technology's most sophisticated use cases, its lighting simulation skills must be given a lot of weight.
They pointed out that the XR system must keep the natural and virtual lighting conditions closely following one another throughout time.
The number of light sources should not exceed the number of hardware-accelerated lights, as rendering must be done in real-time.
It is evident that real-time rendering is impossible for the tens or even hundreds of fixtures provided in a lighting arrangement.
It is necessary to offer suitable techniques that significantly lower the number of light sources used while minimizing the loss of visual quality~\cite{970535}.

Photo-realistic rendering is also important for XR rendering.
It involves the rendering of the scene using an inverse light transport equation.
It has been also known that the least square is able to solve the inverse photo-realistic rendering effectively~\cite{aittala2010inverse}.
While working well in a controlled environment, the practical rendering can be challenging for practical applications.
The variation of outdoor lighting intensity makes it difficult for XR app designers to design one app to be suitable for all these variations.
As such, an adaptive lighting model is needed  under this condition~\cite{behzadan2005visualization,cirulis20133d,pascoal2017information}.
Second, lighting model will be also useful for the XR applications in dark environments such as the applications used at night~\cite{blom2018impact}. 

It is known that XR technology relies on a good amount of lighting to generate high-precision spatial mapping.
Poor lighting makes it difficult for the generation of high-precision spatial mapping~\cite{blom2018impact}.
Without a correct spatial mapping, it is infeasible to correctly place virtual objects into the scenes, therefore, making it hard to use XR app under such conditions.
To resolve this problem, a scene illuminator needs to be used.
Presently, Hololens is equipped with scene illuminators for this purpose.
The design of scene illuminators is however challenging considering the variation of lighting conditions.
The design of scene illuminators can be made better produced through the use of light simulation software.
For designing such illuminators, lighting simulation software can be used such as Radiosity and Photon Mapping.

Conversely, XR would not work well if there is too much light either.
This would occur when using the XR directly toward sunlight.
Under this condition, sunlight flux can become a major issue~\cite{pascoal2018survey}.
Most of XR device manufacturers have clearly stated that the use of XR needs to be cautious while using the device under direct sunlight~\cite{dorreneb_2022_hololens}.
However, it is not rare to find such applications in practice, e.g., construction, farmers, and welding industries.
Therefore simulation of the light intensity and how it affects XR virtual object rendering needs to be well-addressed.
For simulation of the adverse impacts of strong light flux on XR virtual object rendering, high-level lighting simulation is not helpful because they are mostly designed as either view or scene-based lighting simulation.
To resolve this problem, low-level lighting simulation software is needed.

Low-level lighting simulation for XR device simulates the light propagation of XR device~\cite{yang2021fast, yang2020high, yang2020high1, yang2019fast, yang2018large}.
The light propagation of the XR device follows the light diffraction rule.
Light diffracts in the air consisting of near-field diffraction (Fresnel diffraction) and far-field diffraction (Fraunhofer diffraction).
XR device follows the near-field diffraction rules.
To manipulate the projected virtual object intensity and counter the effect of sunlight flux, light intensity from Fresnel diffraction needs to be well controlled.
For this reason, the light source, typically RGB LED diodes, should be carefully chosen to satisfy the needed projection of object intensity.
Furthermore, adaptive light intensity filters are needed to enable adaptive light projection.
The simulation software needs to consider these factors to get the appropriate light intensity in the presence of extreme sunlight flux.
To design XR device filters and light sources for adaptive virtual object lighting intensity, professional ray tracing-based optical device simulation Zemax is needed.
Zemax is a proven software known able to simulate optical devices with high accuracy.
However, Zemax is a general-purpose optical device simulation software.
It means that it is not optimal and time-consuming to use Zemax for the simulation of the XR device.
More specific AR-specific simulation software such as Synopsys LightTools should be considered.
Different from Zemax, Synopsys LightToolscan be used to simulate the grating design of the AR device.
The grating of the AR device is located in the glasses of the AR device.
The grating is a critical optical part to display the images, especially in low or strong lighting conditions.
Light tools is able to simulate different ambient lighting conditions and how it impacts the displayed virtual objects therefore useful to design AR devices.

\subsubsection{Lighting in Virtual Reality}

The use of immersive VR as a lighting design tool is a topic that has piqued the scientific community's curiosity.
Simulating lighting in VR involves the use of advanced graphics engines, computer-generated imagery, and real-time rendering techniques to accurately replicate the behavior of light in a simulated environment.
This is achieved by modeling light sources, materials, and their interactions with the environment to create a realistic and immersive lighting simulation.
Additionally, modern VR systems utilize high-resolution displays and optics to present this simulated environment to the user, further enhancing the sense of realism and immersion.
VR Technology would allow for a comprehensive analysis of lighting alternatives, encompassing factors like aesthetics, environmental light pollution, and energy efficiency.
Additionally, this analysis would be intelligently facilitated through complete computer assistance~\cite{krupinski_virtual_2020}.

The ability to reproduce a light environment that appears to be real, a correct light distribution from a photometrical standpoint, and the impressions that people have in real locations are all important factors in employing VR environments as a substitute for real ones.
VR is emerging as a valid alternative for evaluating the perception of the indoor visual environment due to the ability to control selected variables, analyze cause-effect relationships, and save time and cost~\cite{bellazzi_virtual_2022}.
In addition, immersive VR makes it possible to circumvent some obstacles to conducting experiments in a real-world setting, such as lighting, by modifying the visual environment and securely investigating hazardous environments~\cite{scorpio2020virtual}.
By enabling the control of lighting conditions and rapid alternation of visual stimuli, as well as by providing a high degree of immersion—a factor that has been identified as crucial in recreating the experience of real space—this technology has the potential to overcome significant obstacles in conducting experiments in real environments~\cite{chamilothori_adequacy_2019}.

VR can accurately replicate the lighting characteristics of a physical setting, including open/closed, diffuse/glaring, bright/dim, and noisy/quiet~\cite{chen2019virtual}.
However, another study indicated that the main restriction on the use of VR for lighting research is the lack of a standardized investigation approach, along with the inability to identify the most valued setting parameters (CCT, illuminance, source typology, daylight conditions), as well as to observe people's behavior with regard to their interactions with lighting controls and shading devices~\cite{bellazzi_virtual_2022}.
The study~\cite{scorpio2020virtual} confirmed a research gap in determining how accurately gaming engines reproduce the dispersion of light; as a result, immersive virtual reality might be utilized for light design considerably more.
In~\cite{bellazzi_virtual_2022} comparisons between luminous conditions in VR and real environments has been conducted.
Another study~\cite{keshavarzi_radvr:_2021} introduced RadVR, a VR tool for daylighting analysis that integrates qualitative evaluations simultaneously with immersive real-time rendering to offer potential support for tasks requiring spatial knowledge, navigation, and analysis of the position of the sun in VR.
Study~\cite{may2020vrglare} presented VRGlare, an immersive lighting
performance simulator that incorporates VR to provide real-time glare modeling and analysis.
While VR has shown promise in lighting design, there is still a perceptual divergence between virtual and actual lighting experiences that needs to be addressed.
Future research should focus on improving the accuracy of virtual simulations, exploring different display technologies, and understanding user factors that influence perception.
By narrowing this perceptual gap, VR can become a valuable tool in the field of lighting design, facilitating efficient prototyping and adaptation of lighting designs.

\subsubsection{Lighting in Augmented Reality}

Most of the research has focused on exploring VR in the fields of lighting design and research, while AR has not received an equal level of attention.
AR which refers to any interface that superimposes digital content onto real-world as described in the reality-virtuality continuum~\cite{milgram1994taxonomy}, can be a promising solution for digital twin technology where the real and the virtual world are integrated.

Simulating lighting in AR involves the use of computer vision algorithms and machine learning techniques to detect the physical environment and accurately overlay virtual lighting elements onto real-world surfaces.
By using the camera and other sensors on an AR device, such as a smartphone or tablet, the system can accurately detect the position and orientation of virtual light sources and simulate the behavior of light in the real-world environment.
This allows designers and architects to create and visualize lighting designs in situ, providing a more accurate representation of how lighting will appear in real-world conditions.

The integration of digital twin and AR technologies in lighting design offers several potential benefits, including the ability to create and test lighting designs in a virtual environment before implementing them in the real world.
A digital twin is a virtual replica of a physical object or environment that can be used to simulate and test different scenarios, while AR overlays virtual elements onto the real world, creating an interactive and immersive experience.
By combining these technologies, designers can create a virtual replica of a real-world environment, overlay virtual lighting elements onto it, and use AR technology to visualize and test how the lighting design will appear in the physical space.
This approach can help designers identify potential issues and optimize their lighting designs, leading to more efficient and effective solutions.

Digital twin has emerged for modeling and simulating data and processes, creating a variety of potential for assisting the decision-making process based on the evaluation of realistic scenarios~\cite{Hassan2022DigitalTwin}.
The same study presented a digital twin light modeling for design in which researchers investigate the design, development, and simulation of lighting interventions, as well as of analyzing possible light effects in the urban/outdoor context.
A study by Sheng et.al~\cite{5342415} explored the use of a spatially augmented reality (SAR) sketching interface for architectural daylighting design.
The study focused on the development and evaluation of an SAR system that allows architects to sketch their designs and assess the impact of daylighting in real-time.
The system consists of a custom-built projector and a Microsoft Kinect sensor, which are used to project virtual daylighting scenarios onto physical models of buildings.

Another study~\cite{10.1145/3550291} described LitAR, a system that addresses the spatial variance problem in lighting reconstruction for mobile AR applications.
The system employs two-field lighting reconstruction, two noise-tolerant data capturing policies, and two real-time environment map rendering techniques to handle the computation requirement mismatch.
LitAR offers knobs for developers to make trade-offs between quality and performance.
Evaluation results show that LitAR achieves visually coherent rendering effects, significantly reduces point cloud projection time, and outperforms a recent work in terms of Peak signal-to-noise ratio (PSNR) value.

Similar to VR, the accuracy of AR in emulating real lighting conditions requires careful evaluation to guarantee its dependability.
AR's strengths lie in its capacity for on-the-spot collaboration and decision-making, yet there is a clear need for more research to thoroughly assess its effectiveness and establish optimal practices for its implementation in lighting design.

\subsubsection{Lighting in Mixed Reality}

Simulating lighting in MR involves the use of specialized headsets equipped with cameras and sensors to track the user's environment and integrate virtual objects with the real world.
The addition of spatial mapping and gesture recognition enables the user to interact with virtual lighting elements in a more natural and intuitive way, allowing for a more immersive and responsive experience.
By using these technologies, designers and architects can create and test lighting designs in a VE that accurately reflects real-world conditions.
The study~\cite{366462844} in lighting design presents a method for generating accurate shadows of virtual objects in a mixed reality environment.
Researchers in this study described a method for displaying realistic shadows of virtual objects in a MR environment by reconstructing the light source distribution of a scene in real-time through the segmentation and analysis of a known occluding object's shadows.

Park et al. proposed a method for estimating lighting in mixed reality environments.
The method uses a deep learning algorithm to estimate lighting from a homogeneous-material object, which serves as a proxy for the real environment.
The paper describes the technical details of the method, including the architecture of the deep learning model and the training process.
The method is evaluated using a series of experiments that demonstrate its effectiveness in estimating lighting in a variety of MR scenarios.
The paper concludes that the proposed method is a promising approach to solving the problem of lighting estimation in MR environments, and that it has the potential to significantly improve the quality and realism of MR experiences~\cite{8998303}.

Another study~\cite{chamilothori_adequacy_2019} introduced a method for generating photo-realistic renderings of MR scenes.
The method uses image-based lighting and physically-based rendering techniques to create realistic lighting effects consistent with the real-world environment.
The paper outlines the technical details of the proposed method, including the process for capturing and processing the real-world environment and the algorithm for rendering the mixed reality scene.
Santos et al.~\cite{Santos2007DISPLAYAR} addressed various display technologies, including head-mounted displays, projection systems, and desktop displays, as well as different rendering techniques, such as rasterization, ray tracing, and real-time rendering.
The paper also discusses the challenges associated with displaying and rendering complex 3D models in MR environments, such as the need for high-resolution textures and efficient rendering algorithms.

The fidelity of MR in simulating lighting conditions requires thorough investigation to ensure that it can be trusted for use in design and research.
The advantage of MR lies in its ability to provide an immersive and interactive experience that can be manipulated on the fly, but as with AR and VR, further research is necessary to fully understand its capabilities and to craft guidelines for its application in lighting design.

\subsubsection{Lighting and Evaluation}

Assessing the influence of lighting on the rendering of virtual objects, whether in XR devices, is crucial for designers.
The disparities between VR and AR devices lead to variations in the generated content.
In contrast to VR, AR content is the overlay of the object information on top of real-world objects, therefore the lighting model needs to consider the ambient background information.
The perception of VR-AR generated content can be precisely captured and evaluated using a holistic approach.
Such a holistic evaluation approach attempts to model the scene rather than treating each object evaluation separately, which is known ineffective to capture the actual user perception of the AR.

Peak signal-to-noise ratio (PSNR) and structural similarity index measure (SSIM) are two common measures for the evaluation of VR and AR image quality.
These two measures are able to evaluate the noises of the images and the similarity of the rendered images in contrast to ground truth.
These two measures are known to work well and robust for not only AR image quality evaluation but also applicable to many other fields as well such as microscopy images, image denoise,  super-resolution imaging, and image generation.
To deal with the specialty of VR and AR images, which are omnidirectional images,
Craster parabolic projection PSNR and a weighted-to-spherical PSNR namely CPP-pSNR and WS-PSNR are also introduced through the evaluation process~\cite{zakharchenko2016quality,sun2017weighted}.

Subjective measures are also used for omnidirectional image visual quality assessment (VQA).
Mean opinion score (MOS) and differential Mean opinion score (D-MOS) are used for subjective user evaluation, where later accounts for the differences of consequent images displayed to users~\cite{tran2017evaluation,zhang2018subjective}.
Specifically for the assessment of omnidirectional images, the view angles and their impacts on the opinion score are weighted in, yielding the new measure namely O-MOS.
The O-MOS is formulated by having users watch the baseline (good quality) and impaired quality of the omnidirectional images, following watching, the users are expected to assess the images.
O-MOS value is calculated by weighting the differences of the image quality~\cite{xu2017subjective,xiu2017evaluation}.
Furthermore, the separation of 360 video into different regions and performing the subjective assessment of each such region is also important for detailed omnidirectional image evaluation~\cite{xu2018assessing}.

Deep learning-based evaluation is also used for omnidirectional image VQA.
Early attempts include the Convolutional Neural Network (CNN)-based omnidirectional image salience (attention) deep learning method~\cite{assens2017saltinet,luz2017saliency}.
The CNN-based deep learning while effective is not able to predict image quality, it tends to over-fitting the generation of less confident prediction results.
It is because the high-frequency noise in the high-quality image tends to cause over-fitting problems.
To alleviate such concerns,  a smoothing term is introduced to smooth out the high-frequency noises thus reducing the adverse impacts of the high-frequency noises to reduce the over-fitting problem~\cite{kim2017deep}.
Some large data set has been created to fill the gap of small data set problem since deep learning requires large data set for training and validation, as such it has become more practical to use deep learning to perform  omnidirectional image VQA~\cite{li2018bridge}.

Adversarial (GAN network approach) learning is also used for the VQA  of omnidirectional images.
This work consists of objective VQA and subjective VQA.
The learning-based framework uses the objective VQA to deceive the subjective VQA, therefore, achieving high-quality and accurate objective VQA.
More specifically, the objective VQA uses positional and visual characteristics of the omnidirectional image by encoding the positional feature and visual feature of a patch on the omnidirectional image.
It follows by the deceiving process of the subjective VQA which serves as the guider of the training process.
Results show that the proposed adversarial learning is able to achieve state-of-the-art results outperforming PSNR-based evaluation methods~\cite{kim2019deep}.
Such adversarial learning is not only used for the assessment of image quality but also the salience mapping, which is another important measure of the VQA~\cite{pan2017salgan}.
Furthermore, deep reinforcement learning has been known able to extract the salience map effectively~\cite{xu2018predicting,xu2020state}.

In comparison with VR VQA, AR VQA uses similar measures.
Among them, MOS is an effective measure.
Similar to VR VQA, both reference images and impaired images are needed for image quality assessment.
The difference is that VR VQA uses omnidirectional images, whereas AR VQA requires the use of conFusion images.
The conFusion images are the fusion of both foreground image (virtual object image) and background (real-world) images.
Such conFusion images are similar to the impaired images in VR VQA.
For this,  a large-scale state-of-the-art ConFusing Image Quality Assessment (CFIQA) database has been established.
It has 600 reference images and 300 distorted images generated by mixing reference images in pairs~\cite{duan2022confusing}.
Use of the CNN-based deep learning approach, it has shown that such a large database is able to achieve object assessment score matching with subjective assessment score, therefore, showing the effectiveness of the deep learning-based AR VQA~\cite{duan2022confusing}.
This method is also better than conventional AR VQA, which typically lacks of the background scene and only the foreground image is evaluated~\cite{alexiou2017towards}.

While lack of literature dedicated particularly to how lighting influences the VR and AR VQA, insights related to it still can be offered in the light of aforementioned literature related to VR and AR VQA.
The direct impact of lighting on the results of image quality can be well simulated with scene-based lighting simulation tools such as Radiosity and Photon Map.
The simulated lighting impacts on the user perception can be done through the creation of large-scale simulated VR and AR images using similar procedures of existing data set such as CFIQA~\cite{duan2022confusing}.
Similar deep-learning frameworks such as CNN, adversarial, and reinforcement learning can be implemented to bridge the gap between object evaluation scores with subjective evaluation thus achieving a state-of-the-art lighting-related VR and AR VQA.
It is expected that such kind of work will contribute to the understanding of the impacts of lighting on human behavior and activities stated below.

\subsection{RQ3: What are the effects of lighting on human behavior and activities?}

Modern lighting significantly impacts all aspects of human life as it improves the quality of city users in SBE~\cite{mohammadrezaei2022extended}.
Light is the stimulus that affects individuals' perception, performance, and well-being in everyday life.
Numerous research has looked into how lighting affects people's health~\cite{boyce2022light,Dale_1990_toward,RQ1_Naoshi16ComfortableIndoor,366462927,KNEZ2001Effects}.
The impact of light exposure on human health can be either positive or negative, and direct or indirect~\cite{boyce2022light} depending on variables such as gender, age, culture, and so on.

This section examines light's physical and psychological aspects, investigating the connections between light-based emotions, behaviors, and performance in illuminated environments in various sensory scenarios.
The goal is to compile the available research on the relationships between indoor lighting (both natural and artificial) and overall health, well-being, and productivity.
As a main component of SBE, smart lighting's primary objective is to satisfy inhabitants' needs while promoting user ease, comfort, and security~\cite{soheilian2021smart}.
Therefore, the impact of light on SBE inhabitants requires strong attention.

\subsubsection{Mood and Emotions}

Studies show that certain impressions and emotional reactions in people can be triggered by lighting~\cite{Dale_1990_toward}.
It can impact mood directly by influencing neurotransmitter availability, such as serotonin, which is involved in mood regulation, and indirectly by stabilizing circadian rhythms~\cite{blume_effects_2019}.
The results from some related studies suggest that besides exploring traditional elements of lighting quality, which explains the visual aspect of lighting, new insights about a non-visual response to light are required to examine.
This includes the impact of light on emotional state.

Color temperature and light intensity are the two crucial lighting characteristics that have a unique psychological and cognitive impact on people.
Researchers have looked into how the color of light affects mood.
Knez~\cite{KNEZ2001Effects} examines how individuals' self-reported mood, cognitive function, and perception of room light were affected by morning color (`warm,' `cool,' and `artificial daylight' white illumination).

Since lighting significantly impacts the human condition, it is also used for therapy.
To treat mental diseases, light therapy (LT) has become more and more popular over the past few decades~\cite{schwartz_psychiatry_2015,wirz-justice_chronotherapeutics_2013,blume_effects_2019}.
Although the mechanics underpinning LT require further study, there has been multiple research on this area~\cite{Dale_1990_toward,schwartz_psychiatry_2015}.
Lighting can also help in reducing depression, anxiety, and sleep disorders.
Moreover, growing evidence shows that it can be used to treat dementia, delirium, and attention-deficit/hyperactivity disorder~\cite{schwartz_psychiatry_2015}.
There also exists evidence that chronic fatigue can be controlled by light treatment.
Fatigue can impair some elements of mood, cognitive, psychomotor functioning, and work performance.
According to research~\cite{butler_series_2022}, using both bright and dim light treatment may effectively lessen fatigue symptoms.

Positive technologies may improve social, psychological, or subjective well-being (hedonia), according to some theories~\cite{botella_commentary:_2020,pavic_because_2022}.
VR is one of the technologies mentioned in the context of ``positive technologies,'' particularly as a ``hedonic'' technology that enables joyful experiences in the present.
VR is a good contender for developing positive experiences because of its immersive capacity and the sense of presence that it creates.
There is not a fair amount of research evaluating humans' positive and negative moods in VR for accurate assessment.
According to~\cite{pavic_because_2022}, VR can induce a range of positive emotions, including joy, relaxation, and more complex feelings like awe (i.e., a sense of wonder when faced with vast and transcendent stimuli) and the sublime (i.e., a sense of ``amazement tinged with fear'' in response to powerful or large-scale stimuli). While AR might not be as immersive as VR, its integration into daily life offers a seamless way to provoke positive emotions and possibly even elevate moods.

\subsubsection{Seasonal Affective Disorder}

LT is primarily used to treat seasonal affective disorder (SAD), a type of seasonal depression that is more prevalent in the fall and winter when people are exposed to less sunlight.
The leading cause of SAD is the lack of synthetic hormones in the brain caused by the short daylight hours~\cite{Partonen1998SAD}.
Numerous studies have demonstrated the benefits of full-spectrum illumination on seasonal affective disorder~\cite{hawes_effects_2012}.
SAD treatment using bright light therapy (BLT) was first made available in 1984~\cite{schwartz_psychiatry_2015,rosenthal_seasonal_1984} and since then, a significant number of researchers have explored the impact of LT on SAD~\cite{golden_efficacy_2005,martensson_bright_2015,schwartz_psychiatry_2015, Partonen1998SAD}.

A 3D uniform light field can be used to provide healthy lighting space that can be used for SAD therapy and circadian rhythm~\cite{366462927}.
One of the advantages of this approach is that the 3D uniform light source can treat multiple patients simultaneously, even during walking.
They evaluated the effectiveness of their approach by measuring the optical parameters and comparing the result to reality.
The result indicates that the optical power, color temperature, and divergence angle are very close to reality, meaning that the simulated data could be trusted.

VR technologies are playing an increasingly important role in the diagnostics and treatment of mental disorders.
VR can be a game-changer in the growth of strategies for treating depression because many symptoms of depression are feelings-related.
While ample studies indicate the effectiveness of VR intervention to combat SAD~\cite{lindner_how_2019}, the reliability of VR as a tool for assessing specific environmental aspects like lighting quality  for this population is still up for debate.
While studies on VR's effectiveness in treating SAD suggest its promise, the use of AR and MR for similar purposes remains less explored, especially concerning the assessment of environmental factors like lighting.
Therefore, ongoing research into the reliability and effectiveness of AR and MR in these contexts is crucial to validate their role in mental health treatment strategies.

\subsubsection{Attention Deficit Hyperactivity Disorder}

Studies have shown a correlation between light exposure and attention deficit hyperactivity disorder (ADHD).
ADHD is prevalent among adolescents and is estimated to affect up to \%5 of adults worldwide~\cite{brainsci11101361,PrSaSeInXR}.
High rates of neuropsychiatric comorbidities, such as anxiety and depression disorders, are seen in adults with ADHD~\cite{brainsci11101361,surman_adhd_2013, PrSaSeInXR}.
The usage of BLT in ADHD treatment has been addressed by multiple studies~\cite{rybak_open_2006,kooij_high_2014} complementary treatment for ADHD symptoms in adults.
Morning BLT was associated with improved mood symptoms, a considerable phase advance in circadian preference, and a significant decline in subjective and objective indicators of core attention deficit hyperactivity disorder pathology~\cite{rybak_open_2006}.
Given the current importance of light therapy in treating ADHD, VR can be an excellent tool for simulating morning light and for providing the ADHD population with personalized light exposure based on their diverse conditions.
AR could offer more flexibility and practicality by overlaying therapeutic light settings onto a user's real-world routine, making it a convenient tool for continuous light-based interventions.
MR's unique ability to blend real-world and virtual elements might also enable more nuanced control over the light therapy parameters, offering a tailored approach that can adapt to the dynamic conditions and needs of those with ADHD.
As with AR and VR, however, the application of MR in this context requires thorough research to optimize its effectiveness and usability.

\subsubsection{Effect of Light on Social Life}

Light has always been a stimulus for the emergence of social activity.
We perform a larger variety of activities in the settings we often inhabit.
As a result, lighting needs to be sensitive to adjust to the range of activities being carried out.
Researchers indicate the role of lighting on people's social life, but only a few studies examine the effects of lighting on interpersonal communication~\cite{sommer_tight_1974,sommer_personal_1969, chaikin_effects_1976, gifford_light_1988}.
In~\cite{carr_effects_1974, gifford_light_1988}, researchers represent that dim lighting lessens eye contact and increases conversation latency.
According to the study~\cite{Evensen2014} people engage more with their surroundings in light environments with high light ratios, where socially significant objects are the major focal points.
In contrast, people interact more with each other in low-level lighting.

Multiple efforts have been made to evaluate the impact of XR technologies on social life, but there is not a fair amount of data related to light exposure to XR and its effects on human social life.
Light can increase social interaction in a VE due to inducing positive feelings.
VR has become one of the most well-popular digital social environments where individuals can connect, engage, and socialize in novel and immersive ways.
Based on existing research findings, VR empowers us to get together with people from all over the world and share experiences that aren't possible to have in the actual world.

\subsubsection{Safety Perception}

The correlation between lighting and perception of safety is a complex issue.
Any inhabited environment must have safe circumstances, so it is important to consider how light affects safety.
The environment ought to be planned to compensate for human potential limitations at least partially.
A person is more likely to notice a hazard and take the necessary action if something helps their vision.
When lighting is involved in accidents, in many cases, low illuminance levels or poor lighting quality can be blamed as the root cause.
There may be poor visibility as a result of all of these.
Adequate lighting is required for visual performance and safety and to prevent falls and injuries~\cite{osibona_lighting_2021}.

Lighting significantly impacts people's perceptions of safety after dark, but the mechanisms by which lighting affects perceived safety are largely unexplored.
There has been limited research on indoor lighting design concerning safety concerns, while most studies concentrate on outdoor lighting and perceptions of safety.
In one study, researchers have looked into how humans perceive light, particularly how they perceive safety, depending on the brightness, uniformity, and position of the light~\cite{scorpio2020virtual,nasar_lighting_2017}.
Another study reinforces the notion that artificial light at night exposure may negatively affect psychological functions~\cite{cho_effects_2015}.
Since there is a strong relationship between safety perception and lighting, safety consideration in lighting design is a matter of growing concern.
Recognizing the multifaceted impact of lighting on safety, further exploration of these dynamics is essential to create environments that promote well-being and security.
Little efforts have been made to evaluate the correlation between lighting and user perception in XR technologies.
Although findings are limited, there is an indication that brighter lights and well-lit environments actually make people feel safer in VR.
In addition, VR and AR have the ability to successfully boost safety and security in various industries~\cite{orji_virtual_2022}.

\subsubsection{Task Performance}

Along with the effects on overall human health and well-being, lighting also affects task performance~\cite{boyce2003human}.
Recent studies~\cite{kent_effect_2009,winkler_treatment_2006,srinivasan_melatonin_2006} have revealed that serotonin and melatonin regulation, which are related to how sunlight and light therapy affect mood, are also involved in cognition, which raises the possibility that light may potentially have an impact on cognitive performance.

The important components of lighting that have been shown to affect task performance are illuminance and color temperature.
Various studies show an increase in task performance with higher illuminance.
Hughes and McNelis~\cite{hughes1978lighting} reported improvements in performance with increased illuminance in difficult paperwork of office workers.
Vimalanathan and Babu~\cite{vimalanathan2014effect} identified that a higher illuminance significantly improves neurobehavioral test reaction time and error response.
Smolders et al.~\cite{smolders2012need} showed the effects of illuminance on subjective alertness and vitality, sustained attention in various tasks.
The underlying principle is that illuminance improves vision, enhancing the ability to perceive optical information and increasing task performance~\cite{van2004lighting}.
However, when the illuminance is sufficient for visual performance, the relative increase in performance decreases. 

Another metric of lighting that affects task performance is color.
Kwallek et al.~\cite{kwallek2007work} investigated the influences of interior colors on workers’ productivity and found that the effects can vary among different individuals.
They also reported that the interior colors could have the opposite impact on individuals' long-term and short-term productivity.
Viola et al.~\cite{viola2008blue} found that blue-enriched white light in the workplace improves self-reported alertness and performance.
Recent studies also demonstrated that long-wavelength red light can increase objective and subjective measures of alertness, resulting in increased task performance.
Mills et al.~\cite{mills2007effect} reported improved self-reported concentration and task performance with fluorescent light sources.
Studies also evaluated the combined impact of illuminance and color on task performance. 
Mills et al.~\cite{mills2007effect} showed that higher illuminance and cooler color temperatures enhance concentration levels.
Whereas lower illuminance and warmer color temperatures seem to improve communication and social behavior~\cite{knez1998effects}.
Barkmann et al.~\cite{barkmann2012applicability} showed a positive impact of variable lighting on standardized tests among school students.

Task performance has been a major focus of previous work developing and evaluating XR applications.
It should be noted that performance is highly affected by incorrect occlusion in AR, and this effect outweighed the influence of spatial characteristics.
In contrast, as more sensory inputs are supplied, and simulation fidelity rises, objective and subjective performance measurements in virtual reality environments rise.

\subsubsection{Other Considerations}

Although light treatment has been shown to be helpful, not everyone will benefit from it.
A more strong, bright, and more precisely focused light might be required for some patients.
The intense light may be too much for some people to take, and it may also worsen some diseases by causing new symptoms to emerge.
Eye damage is also a possibility, but one that happens seldom~\cite{makary_seasonal_2021}.
Another study reveals that many adult outpatients with ADHD report an excessive sensitivity to light~\cite{kooij_high_2014}.
The same study indicated that people with ADHD symptoms wore their sunglasses more during the day in all seasons.
This is associated with Photophobia which is related to the functioning of the eyes, which mediate dopamine and melatonin production systems in the eye.

In~\cite{boyce2022light}, there is an indication that light exposure influences human health positively or negatively.
Improper lighting setup may cause visual discomforts such as eyestrain and headaches or cause damage to the eye and skin.
However, other factors, such as stress, anxiety, and isolation are likely more influential than lighting exposure.
The severity of the effect of lighting exposure can vary from short-lived and insignificant to long-term and fatal, and the same lighting conditions can have different impacts on different individuals. 
Gender, age, personality, psychological health, medical conditions, and other human factors can cause a variation in the impact of lighting.

A significant influence of gender on various room light estimate aspects revealed that females saw the room light, across all light settings, as more expressive than males.
Short-term memory and problem-solving performance were mainly affected by light hue, with individuals doing better in `warm' than `cold' and `artificial daylight' white illumination.
Men fared best in `warm' and `cool' white lighting, whereas women did better than men in `artificial daylight' white lighting.
Gender also influences how social safety is experienced, according to the research.
In fact, it was discovered that gender was one of the best predictors of perceived social safety, with women typically feeling lower levels of social safety in the same lighting conditions as men~\cite{boomsma_feeling_2014}.

The measurement method also needs to be considered when considering the impact of light exposure on SBE inhabitants.
Current subjective measurement tools used for measuring the psychological effect of lighting require re-evaluation and improvement~\cite{Dale_1990_toward}.
State-of-the-art approaches are required to achieve an accurate and comprehensive specification of the luminous environment.
Techniques and analysis used in the past need to be combined into ``a single experimental protocol,'' and newer techniques need to be supplemented to successfully relate the psychological and behavioral effects of lighting.

Immersive virtual environments offer novel opportunities to realistically experience various lighting design cases.
With the advancement in XR technologies color and illuminance levels could be accurately reproduced to simulate lighting conditions.
It can also be an effective tool to evaluate the effect of lighting on visual perception and task performance~\cite{ma2022effects}.

\section{Discussion}
\label{sec:discussion}

\begin{figure}[t]
\centering
\includegraphics[width=0.925\linewidth]{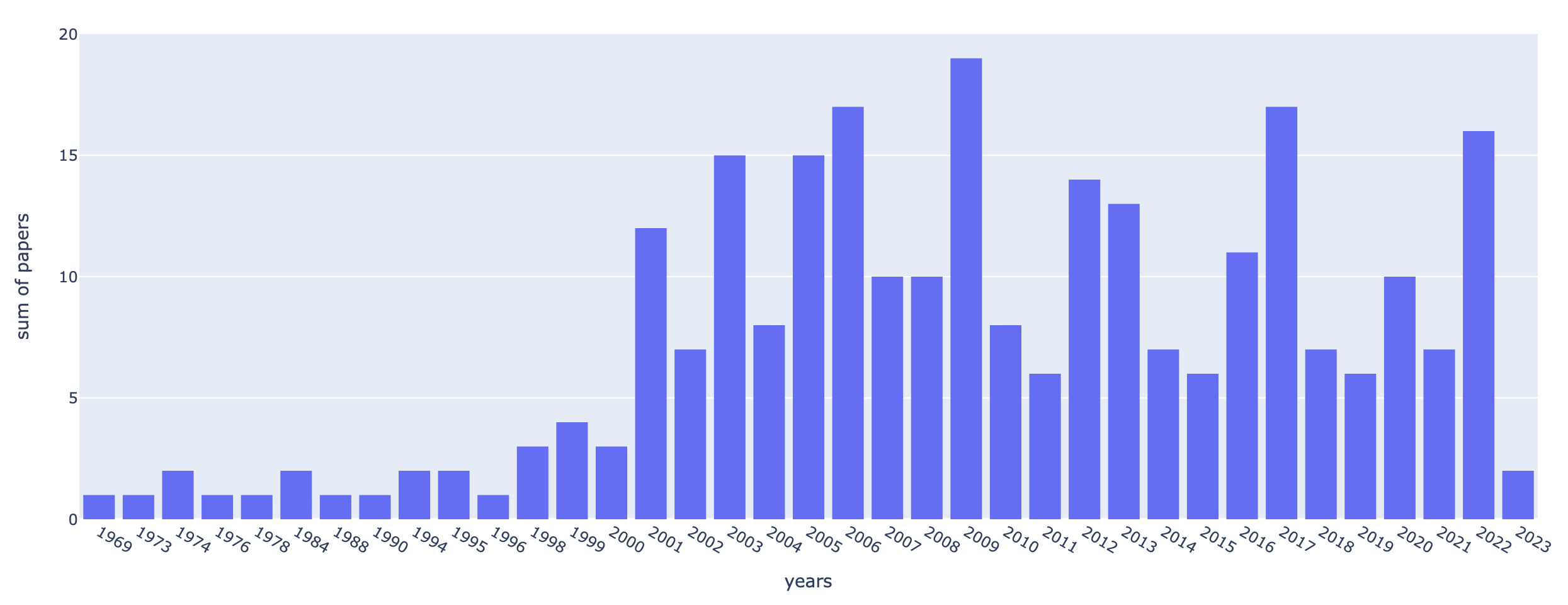}
\caption{Publication timeline (\papercounttotal).}
\label{fig:Year}
\end{figure}

We conducted a systematic review of total of 270 papers to investigate current advancements in lighting simulation tools along with identifying new solutions to satisfy the needs of modern lighting design features.
Figure~\ref{fig:Year} demonstrates the frequency of the papers per year.
Based on reviewing the 65 number of papers which comprehensively compared different tools and algorithms (apart from the remaining 230 papers that scrutinized simulation techniques), in turns out with modern lighting design growth and its complexities, it's hard to benefit from existing simulation tools to achieve an adequate analysis of all the features and opportunities that modern lighting provides.
So, a need for new tools and approaches is inevitable to fill the current gaps in the area of artificial lighting simulation.
To investigate existing solutions we reviewed 41 papers on extended reality technology and solutions it offers.
Based on the potential of XR technology to enhance user experience, a comprehensive analysis of existing literature has revealed a wealth of information pertaining to various lighting design techniques used in XR.
Nevertheless, this systematic review aims to provide a thorough investigation into the role of smart lighting design in extended reality, with a specific focus on the simulation and design of lighting conditions.
The primary objective of this research is to inspire other scholars to delve deeper into the intersection of these two domains, in order to discover new avenues for innovation and design.

On the other side of this study, we addressed reviewing 48 papers to investigate the psychological and physiological effects of lighting design strategies on users.
Although the effectiveness of light and its characteristics like color, intensity and temperature on users' mental and physical health is deniable, there is currently no consensus on how to quantify the stimulus for the psychological effects of lighting.
Our current understanding indicate that people spend more time inside and are more likely to experience depressive symptoms, feelings of loneliness, and isolation during the long, gloomy winters, which makes them less healthy and sociable and more likely to suffer from mental diseases like sadness and loneliness.
Although the effects of lighting on our mood and behavior have not yet been thoroughly established, experts in the field advocate for a shift in the lighting sector towards a greater focus on human health.
The majority of research on interior lighting has examined workplace settings and concentrated on the visual qualities of the area to increase worker happiness and productivity.
There aren't many research that have used real-world trials to expose individuals to various lighting conditions and track how mood and behavior alter in relation to home interiors.
The review sheds light on conducting a systematic review of the literature and synthesize the available data regarding the relationships between indoor lighting, including both natural and artificial light as well as nighttime lighting, and a variety of health outcomes.
To obtain data on various aspects of indoor lighting and its impact on health, including lack of light, various types of illumination, and light as a possible source of hazards, such as from indoor air pollution and burns, light is defined in its broadest sense.
This makes it possible to pinpoint areas that require additional study and their consequences for developing policies.

SBEs are spaces containing sensors, actuators, and intelligent control algorithms. Occupant behavior is complicated because it has three components: social, psychological, and mental well-being~\cite{kumar2019discerning}.
Derek Clements-Croome defines human well-being as \emph{``a concept related to human enjoyment.''}
Furthermore, the author says that human well-being is a measure of a person's quality of life, which meets their physical, psychological, creative, and cognitive demands~\cite{clements2006creating}.
Furthermore, according to Steemers and Manchanda, well-being is a multi-faceted trait that includes health, comfort, and happiness~\cite{steemers2010energy}.
As a result, in a creative setting, well-being can be defined as the underlying feature of productivity levels~\cite{clements2015creative}.

When carefully structured and effectively used, emerging intelligent environment is thought to provide significant opportunities to positively impact human life, both on an individual and societal level, and in particular to provide useful means to support people in their daily activities and thus improve well-being~\cite{stephanidis2019seven}.

It is clear that applications in intelligent environments have the potential to benefit everyone, and this will hopefully promote their adoption in the mainstream market.
However, they are expected to be particularly effective for the elderly and those with physical limitations.
This broad and diverse user population, with a wide range of physical, sensory, and cognitive skills, can benefit from technology applications that help them maintain or improve their health, well-being, and independence, but they may face significant challenges in learning new technologies~\cite{emiliani2005universal}.

There is a fair amount of literature investigating on modern lighting design capabilities.
Despite all the potential use cases and applications for smart cities, energy efficiency remains a major challenge~\cite{366462885}.
Utilizing a proper smart lighting solution leads to reduction in budget and energy.
The integration of the use of LED lights with intelligent management system lead to reduction in urban lighting budget.
Another concern is access to more reliable environmental information which provides better run-time flexibility opportunities.

It is also imperative to explore innovative approaches that can augment the design process and enhance the user experience.
One such avenue is the integration of Large Language Models (LLMs), like Chat GPT, which have emerged as powerful tools for generating creative and effective solutions across various domains.
LLMs are capable of processing and generating natural language text based on the input provided to them, and their potential goes far beyond text generation alone~\cite{kasneci2023chatgpt}.

The versatility of LLMs makes them promising candidates for integration with lighting simulation tools and design processes.
These models can offer a range of capabilities, including design solution recommendations, user-friendly access, iterative design and feedback, and data-driven insights~\cite{ahn2023alternative}.
LLMs can assist designers by providing recommendations for lighting design solutions based on project specifications, user preferences, and energy efficiency goals.
They can analyze complex design parameters and generate suggestions that align with the project's objectives.
They can serve as user-friendly interfaces for lighting simulation tools.
This accessibility can extend beyond professionals to empower ordinary individuals, enabling them to explore lighting design options for their homes, offices, or personal spaces and this can enhance user satisfaction and engagement.
LLMs can facilitate an iterative design process by quickly generating and assessing multiple design variations.
Designers and users can interact with the model to explore different lighting scenarios and receive instant feedback, which can significantly expedite the decision-making process.
LLMs also can process vast amounts of data and provide valuable insights into lighting design trends, energy consumption patterns, and user preferences.
This information can inform the design process and help create more efficient and user-centric lighting solutions.

By incorporating LLMs into the field of lighting design, we have an opportunity to revolutionize the way we approach lighting solutions.
This technology not only benefits professionals in the field but also empowers individuals to take control of their lighting environment.
Moreover, it enables a data-driven approach to lighting design that aligns with modern smart city initiatives and energy efficiency goals.

The integration of AI technologies into smart lighting design has the potential to revolutionize the field by enhancing functionality, adaptability, and energy efficiency~\cite{de_paz_intelligent_2016}.
The applications of AI in smart lighting systems are diverse and encompass adaptive lighting control, predictive maintenance, human-centric lighting, gesture and voice control, energy management, occupancy sensing, personalized lighting profiles, data analytics, predictive design, security and safety integration, sustainability efforts, IoT integration~\cite{9762311}, user feedback analysis, and creative lighting displays.
These advancements will not only improve user experiences and aesthetics but also contribute to sustainability efforts in diverse environments.
However, challenges such as data privacy, algorithm transparency, and system reliability need to be addressed to fully realize the potential of AI in smart lighting design.
Further research and development in this area are necessary to explore the full capabilities and implications of AI-driven lighting systems.

\section{Conclusion}
\label{sec:conclusion}

The evolution of lighting systems design is gradually transitioning into a new stage, where lighting in our environment may be more fully controlled by us and have capability to be integrated into our surroundings.
Throughout the years, our understanding of lighting and its impact on daily life has grown significantly.
Within each step, the importance of studies related to this topic became more evident.
While the traditional approach toward studying the effects of lighting mainly focused on the visual aspect of this phenomenon, new studies suggest that the non-visual aspect also requires attention.

New tools and approaches are required to include multiple aspects of lighting in the study.
In recent years, the advancement of extended reality introduced a unique opportunity to explore lighting aspects that were impossible for researchers to study.
For this purpose, we conducted a systematic literature review to explore recent advances in extended reality technology for smart built environments, particularly for smart lighting systems design.

\section*{Acknowledgment}

Parts of this work have been supported by Virginia Tech Institute for Creativity, Arts, and Technology.

\bibliographystyle{unsrt}
\bibliography{access}

\begin{IEEEbiography}[{\includegraphics[width=1in,height=1.25in,clip,keepaspectratio]{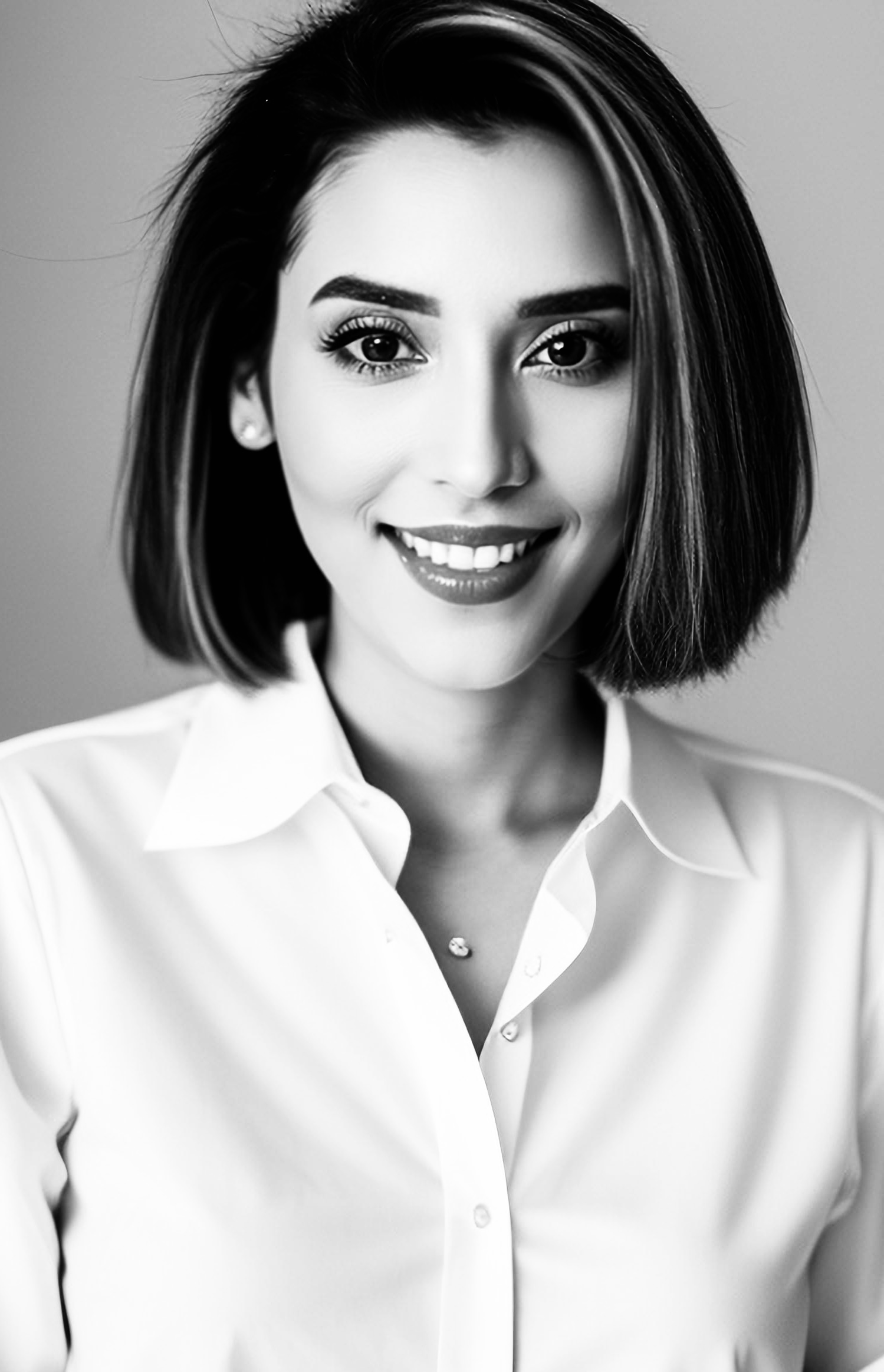}}]{Elham Mohammadrezaei} is a third-year Computer Science Ph.D. student at Virginia Tech.
She has a background in Architecture and received her Master's degree in Architecture from Iowa State University.
Her current research interest stands in the intersection of Human-Computer Interaction, Augmented/Virtual Reality, and User Experience/User Interface design and research, as well as machine learning.
Her e-mail address is elliemh@vt.edu. 
\end{IEEEbiography}

\begin{IEEEbiography}[{\includegraphics[width=1in,height=1.25in,clip,keepaspectratio]{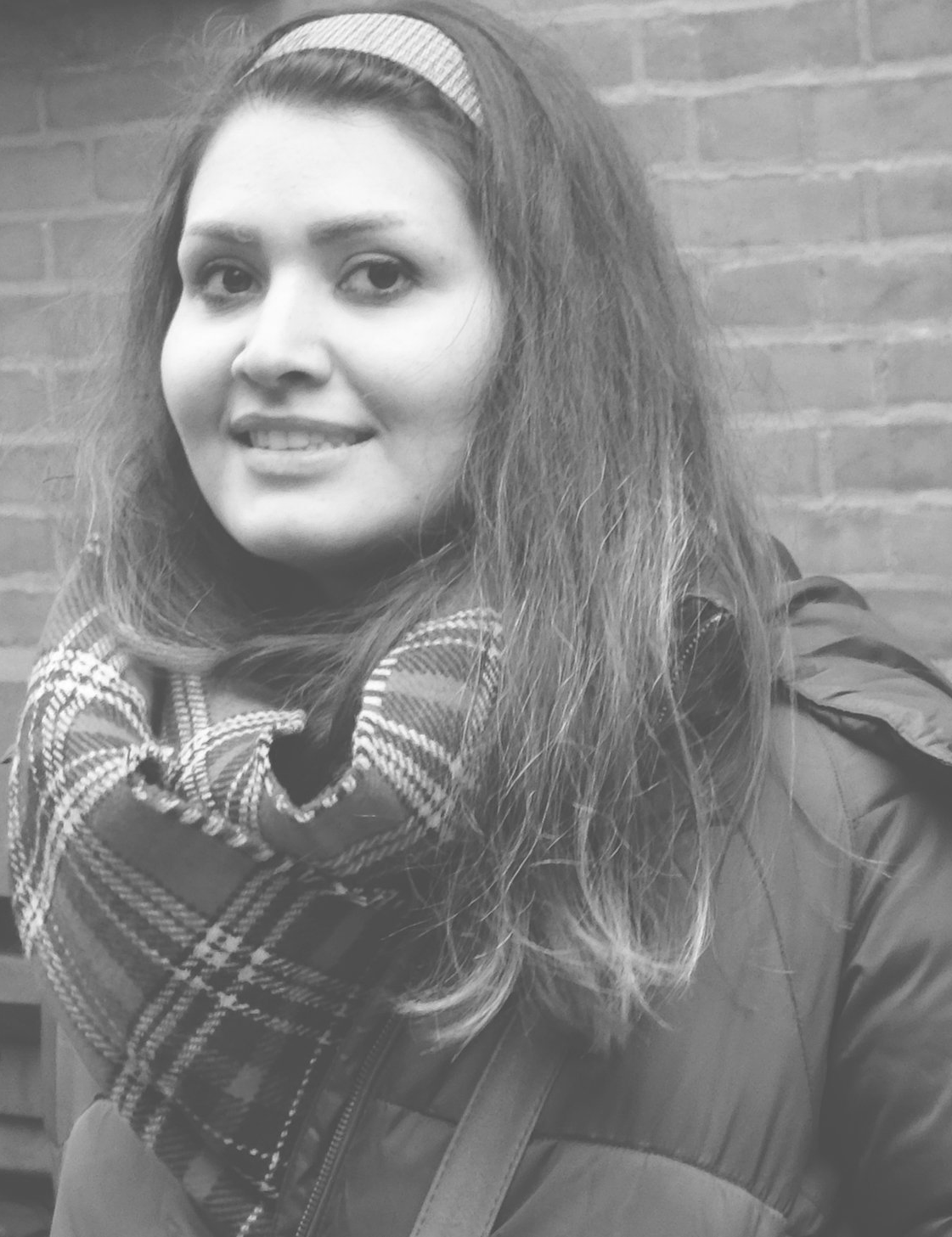}}]{SHIVA GHASEMI} is a second- year Ph.D. student in Computer Science at Virginia Tech.
She received her Master's degree in Human-Computer Interaction from the University of Maryland.
Her research interests include Human-Computer Interaction, Accessibility, and Human Information Processing in augmented reality.
Her e-mail address is shivagh@vt.edu. 
\end{IEEEbiography}

\begin{IEEEbiography}[{\includegraphics[width=1in,height=1.25in,clip,keepaspectratio]{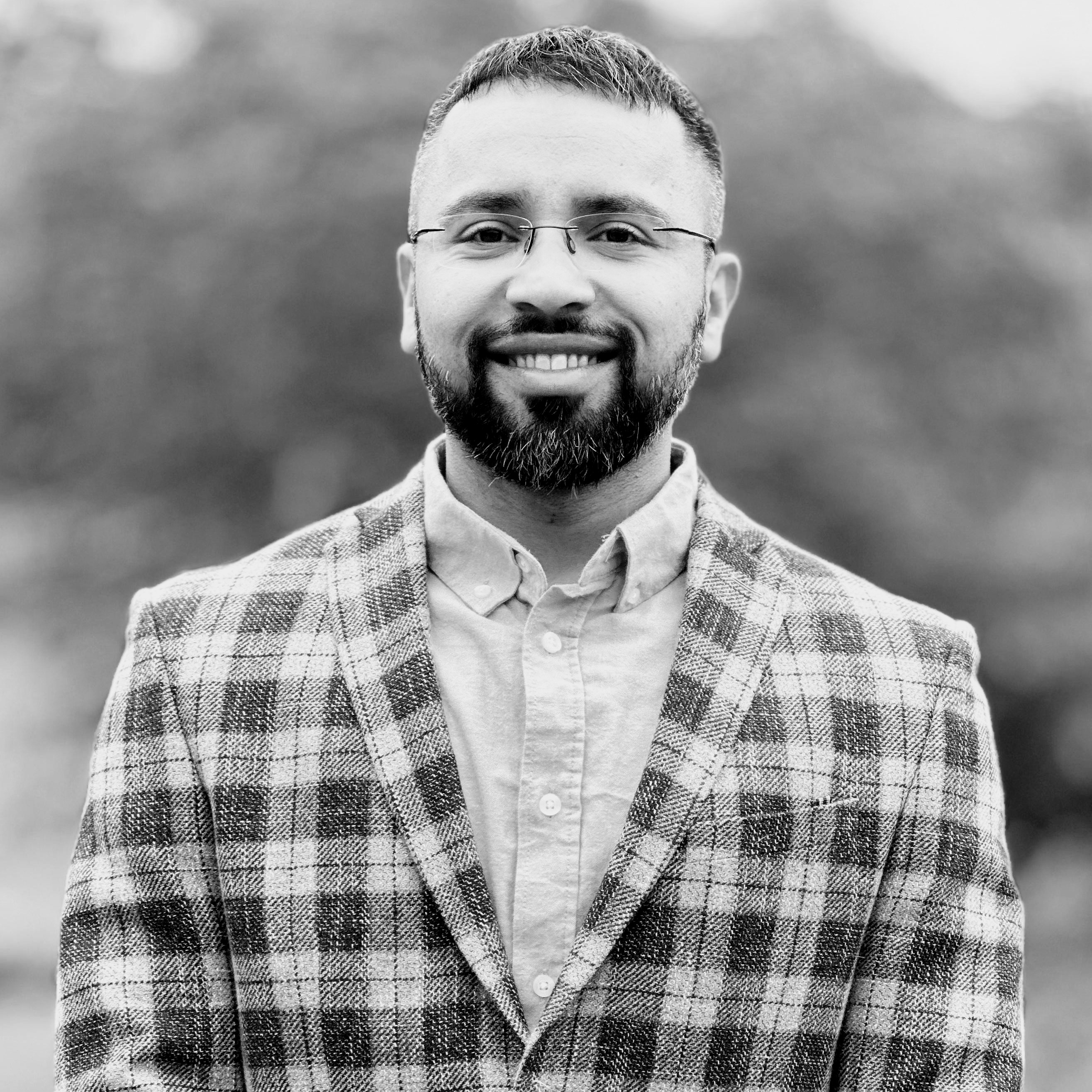}}]{POORVESH DONGRE} is a third-year Ph.D. student in Computer Science at Virginia Tech.
He received his B.S. degree in Civil Engineering from Savitribai Phule Pune University in 2012 and M.S. degree in Construction Engineering and Management from the Indian Institute of Technology Delhi in 2006.
His research interest include Human Behavior Modeling, Machine Learning, and Human-AI Interaction. 
\end{IEEEbiography}

\begin{IEEEbiography}[{\includegraphics[width=1in,height=1.25in,clip,keepaspectratio]{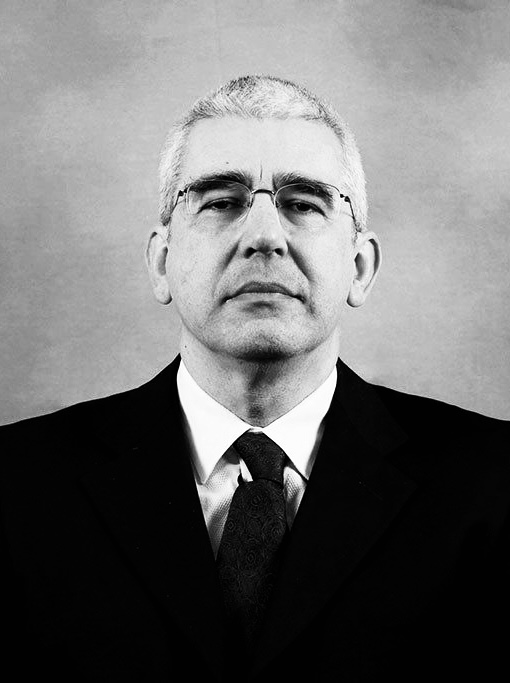}}]{DENIS GRA{\v{C}}ANIN} received the BS and MS degrees in Electrical Engineering from the University of Zagreb, Croatia, in 1985 and 1988, respectively, and the MS and Ph.D. degrees in Computer Science from the University of Louisiana at Lafayette in 1992 and 1994, respectively.
He is an Associate Professor in the Department of Computer Science at Virginia Tech.
His research interests include virtual reality and distributed simulation.
He is a senior member of ACM and IEEE and a member of AAAI, APS, ASEE, and SIAM.
His email address is: gracanin@vt.edu.

\end{IEEEbiography}

\begin{IEEEbiography}[{\includegraphics[width=1in,height=1.25in,clip,keepaspectratio]{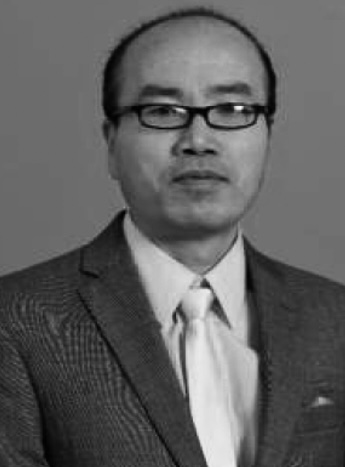}}]{HONGBO ZHANG} received his Ph.D. from Industrial Engineering at Virginia Tech.
He was an Assistant professor in the Department of Computer Science at Virginia Military Institute.
He is currently an Assistant Professor at Mechatronics Engineering of Middle Tennessee State University.
His research interests include Computer Vision, Robotics, and Human-Computer Interaction.
He is a member of IEEE and OSA.
He has served as the guest editor for the Journal of Electronics and Journal of Symmetry.
\end{IEEEbiography}

\EOD

\end{document}